\documentclass{aa}  
\usepackage{natbib}
\bibpunct{(}{)}{;}{a}{}{,}

\usepackage{graphicx}
\usepackage{natbib}

\begin{document}

\authorrunning{I. Ermolli et al.}
\titlerunning{Radiative emission of solar features in the Ca~II~K line}
   \title{Radiative emission of solar features in the Ca~II~K line:
   comparison of measurements and models}
   
   \author{I.~Ermolli\inst{1}, S.~Criscuoli\inst{1},  H.~Uitenbroek\inst{2}, F.~Giorgi\inst{1}, M.~P.~Rast\inst{3} \and S.~K.~Solanki\inst{4,5}}

   \offprints{Ilaria Ermolli \email{ermolli@oaroma.inaf.it}}

    \institute{INAF-Osservatorio Astronomico di Roma, Via Frascati 33, 00040 Monte Porzio Catone, Italy
\and 
   National Solar Observatory, Sacramento Peak, P.O. Box 62, Sunspot, NM 88349, USA
 \and 
Laboratory for Atmospheric and Space Physics, Department of Astrophysical and Planetary Sciences, University of Colorado, Boulder, CO, 80303, USA
\and  
Max-Planck-Institut f\"ur Sonnensystemforshung, Max-Planck-Strasse 2, 37191 Katlenburg-Lindau, Germany
\and  
School of Space Research, Kyung Hee University, Yongin, Gyeonggi 446-71, Korea
}
   \date{}

   \abstract
{The intensity of the Ca~II~K resonance line observed with spectrographs and Lyot-type filters has  long 
served as a diagnostic of the solar chromosphere. 
However, the literature contains a relative lack of photometric measurements of solar features observed 
at this spectral range. 
}
{We study the radiative emission of various types of solar features, such as quiet Sun, 
enhanced network, plage, and bright plage regions, identified on 
filtergrams
taken in the Ca~II~K line.} 
{We analysed full-disk images obtained with the PSPT, by using three 
interference filters that sample the Ca~II~K line 
with different bandpasses. 
We studied 
the dependence of the radiative emission of disk features on the filter bandpass. 
We also performed 
a non-local thermal equilibrium (NLTE)
spectral  synthesis 
of the Ca~II~K line integrated over 
the bandpass of PSPT filters.  
The synthesis was carried out by utilizing the partial frequency redistribution (PRD) with      
the most recent set of semi-empirical atmosphere models  in the literature and 
 some earlier atmosphere models. As the studied models were computed by assuming 
 the complete redistribution formalism (CRD), we also performed simulations 
 with this approximation for comparison.} 
{We measured the center-to-limb variation of intensity values 
for various solar features identified on PSPT images and compared  the results 
obtained with those derived from the synthesis. 
  We find that CRD calculations derived using the most recent quiet Sun model, on average,  reproduce the measured values of the 
 quiet Sun regions slightly 
more accurately than PRD computations with the same model. 
 This may reflect that the utilized atmospheric model was computed assuming CRD.
Calculations with PRD on 
 earlier  quiet Sun model atmospheres reproduce measured quantities 
 with a similar accuracy as to that achieved here by applying CRD to the recent model. 
We also find that  the median contrast values measured for most of the identified bright
features, disk positions, and filter 
bandpasses are, on average, a factor $\approx$ 1.9 lower than those   
derived from  PRD simulations performed using 
the recent bright feature models. The discrepancy  
between 
measured and modeled values decreases by $\approx$ 12\% after taking into account 
straylight effects on PSPT images.  When moving towards the limb, PRD computations display closer agreement with the data 
than performed in 
CRD.  
Moreover, 
 PRD computations on either the most recent or the earlier atmosphere models of bright features 
reproduce measurements 
from plage and bright plage regions with a similar accuracy.
} 
{}


\maketitle

\section{Introduction}
Observations taken with  spectrographs and Lyot-type filters in the Ca~II~K resonance line have long 
served 
as  
diagnostics of the solar chromosphere \citep[e.g.,][ and references therein]{rutten2007} and  
have   been used to determine the 
temperature structure of semi-empirical atmosphere models \citep{vernazza1981,maltby1986,fontenla1991,fontenla1993}.  
However,  except for the quiet Sun \citep{zirker_1968,white_suemoto1968,wittmann1976,livingston2008,grigoryeva2009}, there is a shortage of measurements in 
the literature of the radiative emission 
of  solar  features observed in the Ca~II~K range at various heliocentric angles.  

In addition to providing new data, photometric 
measurements of disk features in Ca~II~K  allow us to check the various semi-empirical models used in  
spectral line computations 
\citep[e.g.][]{rezaei2008,reardon2009,pietarila2009} and the reconstruction of irradiance variations 
\citep[e.g.][]{domingo2009,harder2009,krivova2009}.
In particular, models have been presented \citep[][]{fontenla2009} that 
differ from previous sets of models in terms of the physical assumptions used to derive their temperature structure.
These models have been adapted in particular for irradiance studies. They were tested 
against measurements at EUV/FUV, visible, and infrared wavelengths, but  comparison with 
observations in
the near 
 UV  range  was limited. 
Detailed comparisons between measurements and model predictions in the near UV, including for the Ca~II~K line, would 
aid in  
model refinement. 

In this paper, we present a study of the radiative emission of different types of solar features, such as quiet Sun  (internetwork), network, 
enhanced network, plage, and bright plage regions, identified on full-disk 
observations taken in the 
spectral range of the Ca~II~K line. 
The images were obtained with the PSPT (Precision Solar Photometric Telescope), 
by using 
three interference
filters that  sample the 
Ca~II~K line with different bandpasses. 
We
studied the dependence of the measured emission on the filter bandpass. We also compared the results obtained 
with those derived from the numerical synthesis performed for the bandpass of  PSPT filters  
with the RH code \citep[][]{uitenbroek2002}   
utilizing the new  
 models presented by \citet[][]{fontenla2009}   as well as
earlier atmosphere models \citep{vernazza1981,fontenla1993,fontenla2006}.

\section{Observations and data reduction}
\begin{figure} 
\centering{\includegraphics[width=8.5cm]{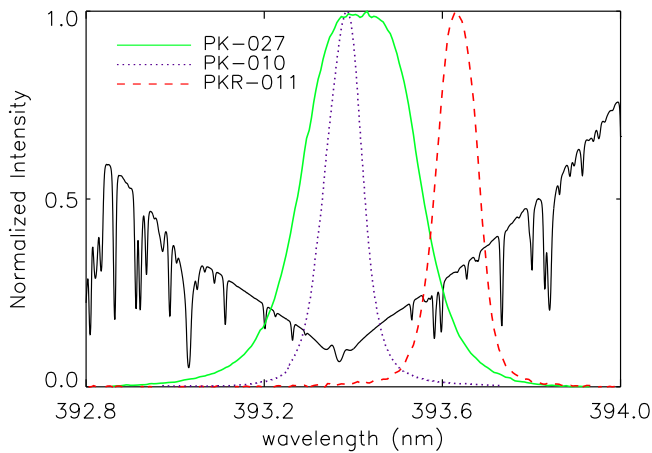}}
\caption{Transmission profiles of the three PSPT interference filters 
 considered 
in this study. The three profiles, each normalized to its maximum, are represented with different colors and lines as indicated in the legend. They are
superimposed on the solar reference spectrum (black line) provided by \citet{delbouille1972}.} 
\label{fig2} 
\end{figure} 

The PSPT  is a 15 cm, low-scattered-light, refracting telescope designed  for synoptic photometric solar observations characterized by  0.1\% pixel-to-pixel 
relative photometric precision 
\citep{coulter1994}. 
Two such telescopes, which differ only in minor hardware characteristics and operational strategies, are operated at  
Mauna Loa (MLSO-PSPT) and 
Monte Porzio 
Catone  (Rome-PSPT) Observatories by the High Altitude Observatory \citep{rast_1999} and Osservatorio Astronomico di Roma  \citep{ermolli1998}, respectively. 
These telescopes 
typically acquire full-disk solar images on 2048$\times 2048$ CCD arrays with narrow-band  
interference filters centred on the
blue continuum (409.412 FWHM 0.267 nm), red continuum (607.095 FWHM 0.458 nm), and Ca~II~K (393.415 FWHM 0.273 nm, hereafter  {\it PK-027}).
However, the installation of additional filters on both the telescopes has also allowed the acquisition of images at other spectral 
ranges. 
In particular, 
two interference filters sampling the Ca~II~K range were added to the MLSO 
telescope in summer 2007, 
one at line center (393.379 FWHM 0.103 nm, hereafter  {\it PK-010}) and the other in the red wing of the line (393.633 FWHM 0.106 nm, hereafter  {\it PKR-011}).
These two filters were operated simultaneously with the broader band  {\it PK-027} filter. Here we analyse 
the  observations carried out with  these three Ca~II~K filters, 
 hereafter referred to as the {\it PK} filters.  

 Figure \ref{fig2} shows 
the transmission profiles measured for the {\it PK} filters
superimposed on a reference solar spectrum. The central wavelength and FWHM of these profiles are summarized in Table \ref{tab1}. 
The filter bandwidths of {\it PK} do not allow the features of the Ca~II~K line center to be resolved. These features, the core, 
the reversal (emission peaks), and the 
secondary minima occur within a spectral range less than 0.1 nm wide. 

We analysed 157 sets of observations available from the MLSO PSPT archive\footnote{Calibrated images from the 
synoptic observations taken with the two PSPTs over many years 
 are available on-line at  http://www.oa-roma.inaf.it/solare and http://lasp.colorado.edu/pspt$_{-}$access/  .} 
for the period 2007 June 7 to July 31. Each set is composed of the images obtained 
with the  {\it PK} filters and one in the red continuum. The red continuum images are, on average,  separated in time  from
the {\it PK}  ones by 7-8 minutes.

 We separated the data into two samples, corresponding to the highest and the lowest quality  images. They contain 37 
and 33 sets of images, 
respectively.
Image quality was determined by measuring the width of the solar limb in the {\it PK-027} images, which provides 
an estimate of the atmospheric seeing conditions 
\citep[][]{criscuoli2007,rast2008}.
In particular, images with an average limb width smaller than 4 pixels  constitute the highest quality  
(hereafter referred to as $good$) images, while the ones with average limb width  larger than 5.5 pixels form the set of 
the worst (hereafter referred to as $poor$) 
images. The average and the standard deviation $\sigma$ of limb width values measured for the whole sample of 
{\it PK-027} images are 4.7 $\pm$ 0.8 pixels; the thresholds introduced above are the values derived from the average $\pm \sigma$. 


All the images analysed for our study were pre-processed  to apply the correction for gain variations across  the 
CCD device and to compensate for residual large-scale linear gradients  affecting the  images after CCD calibration 
\citep{rast2008}.   
The images were re-sized and aligned, 
by using linear interpolation, to a common reference grid on 
which the solar disk size and orientation are constant. 

During the period analysed, the Sun was in a quiet phase. However, 
eight small active regions 
appeared on the solar disk, specifically AR 10958 to AR 10965, in addition to  
some plages and enhanced network regions. 
To separate the various disk 
features seen in {\it PK} images, 
we used the image decomposition method described by  \citet{fontenla2009}. 
This method
 utilizes a threshold scheme derived from the partitioning of intensity histograms constructed  from the images 
 as a function of heliocentric angle
 ($\mu$). The scheme assigns structure types based on  the agreement between the intensity measured at each 
image pixel and the one defined  by the histogram partition. In other words, features are defined by their normalized 
intensity as a function of $\mu$, and pixels are identified with particular features based on their observed position and intensity.

\begin{table}
\caption{ Identification classes, solar features, and atmosphere models utilized in this study.
}  
\footnotesize{
\begin{tabular}{lll}
\hline
\hline
Class & Solar Feature & Atmosphere model\\
 \hline
B & quiet Sun & model B,  VAL3-C, FAL3-C,  FA-06\\
D & network & model D,  FAL3-F\\
F & enhanced &		model F\\
  & network & \\
H & plage	 & model H, model F, FAL3-F, FAL3-P\\
P & bright plage	 & model P\\
S & umbra  & -\\
R & penumbra & - \\
\hline
\end{tabular}
\tablefoot{ The table summarizes the classes  utilized for the 
identification of the various solar features, the relation between each class and the corresponding solar feature, as well as the 
relation between solar features and atmosphere models 
considered for the comparisons presented in the following. 
Models B to P indicate the set of atmospheres of Fontenla 
et al. (2009), which 
constitutes the reference set 
in this study. Models VAL3-C, FAL3-C, and FA06-C indicate the atmospheres for quiet Sun regions 
presented by 
Vernazza et al. (1981), and 
Fontenla et al. (1993, 2006), respectively. Models FAL3-F and FAL3-P are the atmospheres 
for plage and bright plage 
regions presented by Fontenla et al. (1993).}
}
\label{tab2}
\end{table}

The decomposition code used in our study assumes seven classes of disk features identified in {\it PK-027} and red continuum co-aligned images. 
Following both the scheme and reference atmospheres 
proposed by \citet{fontenla2009}, these classes are coded by the letters B, D, F, H, P, S, and R. Features B  through P represent  
 the intensity criterion applied to pixels of {\it PK-027} images. They 
localize average median quiet Sun (internetwork, QS hereafter), network (QS network lanes), enhanced network, plage, and bright plage (or facula) regions, respectively. 
Features S and R, which  correspond to umbral and penumbral regions, respectively,  
 are identified 
in red continuum  images. The threshold values utilized in {\it PK-027} images are shown in the top panels of Figs. \ref{fig9} and \ref{fig9b}. 
We note that feature B is called here  quiet Sun (QS),  but in the threshold scheme localizes more broadly  
internetwork regions, which likely contain residual magnetism.  We also note that the meaning of quiet Sun here is quite 
distinct from that often used in describing chromospheric or 
transition region emission, where it is often used to mean the non-coronal hole portion of the Sun. 
Table \ref{tab2} summarizes the classes utilized to identify  the various solar features, 
the relation between each class and the corresponding solar feature, as well as the relation 
between solar features 
and the  
various atmosphere models considered in the comparisons presented in the following.

\begin{figure}[ht!] 
\centering{
\includegraphics[width=3.5cm]{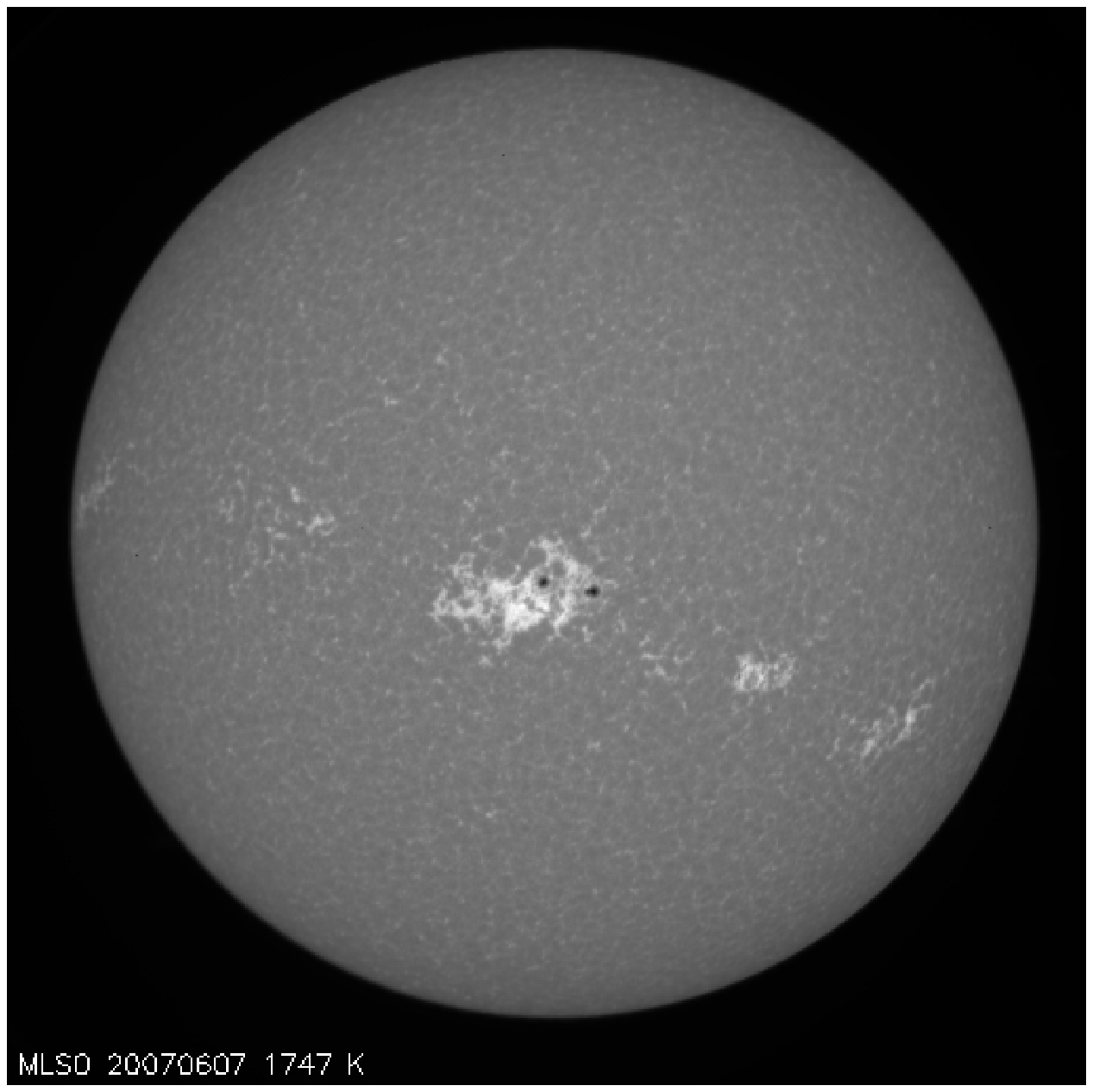}\includegraphics[width=3.5cm]{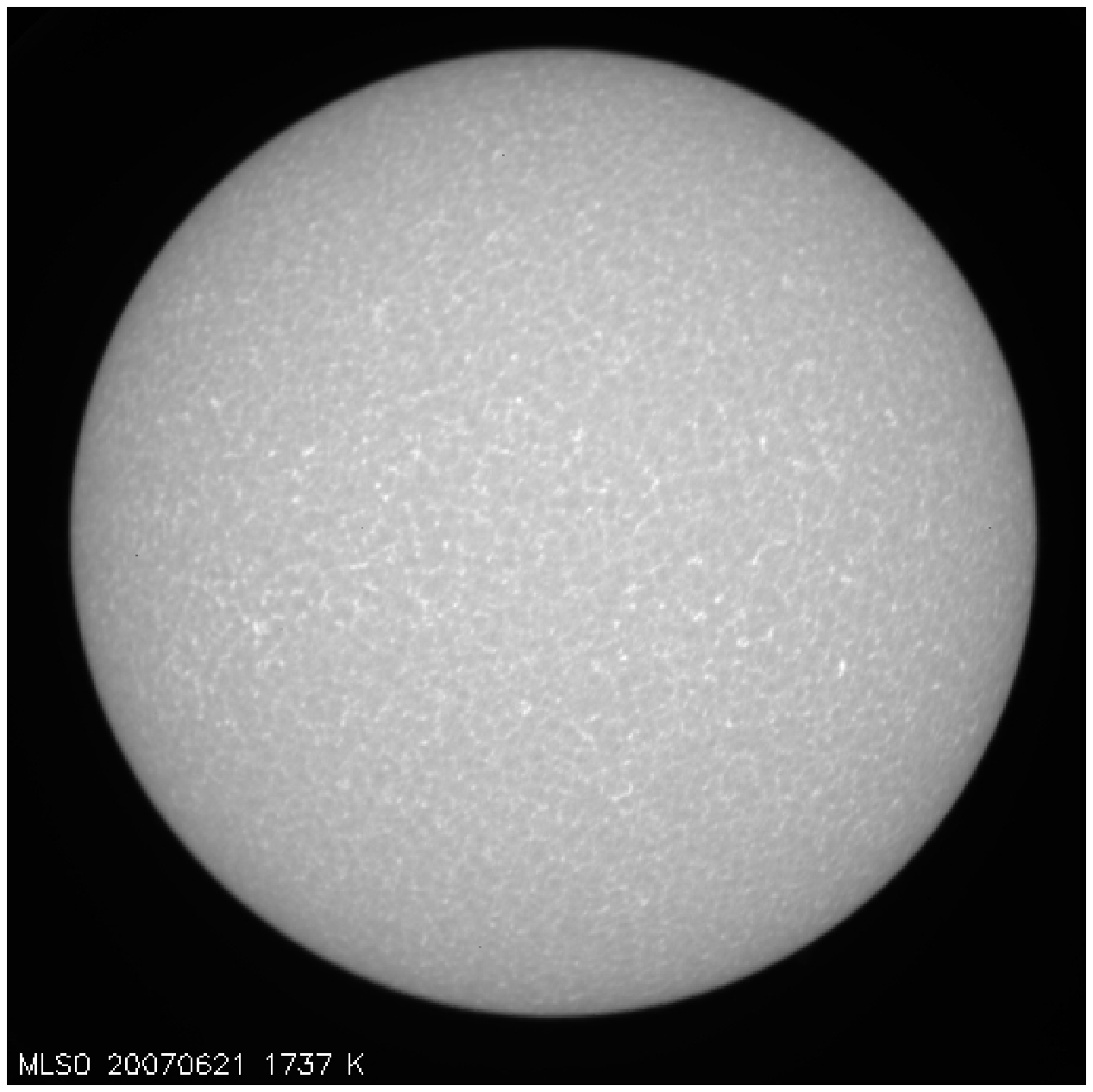}
\includegraphics[width=3.5cm]{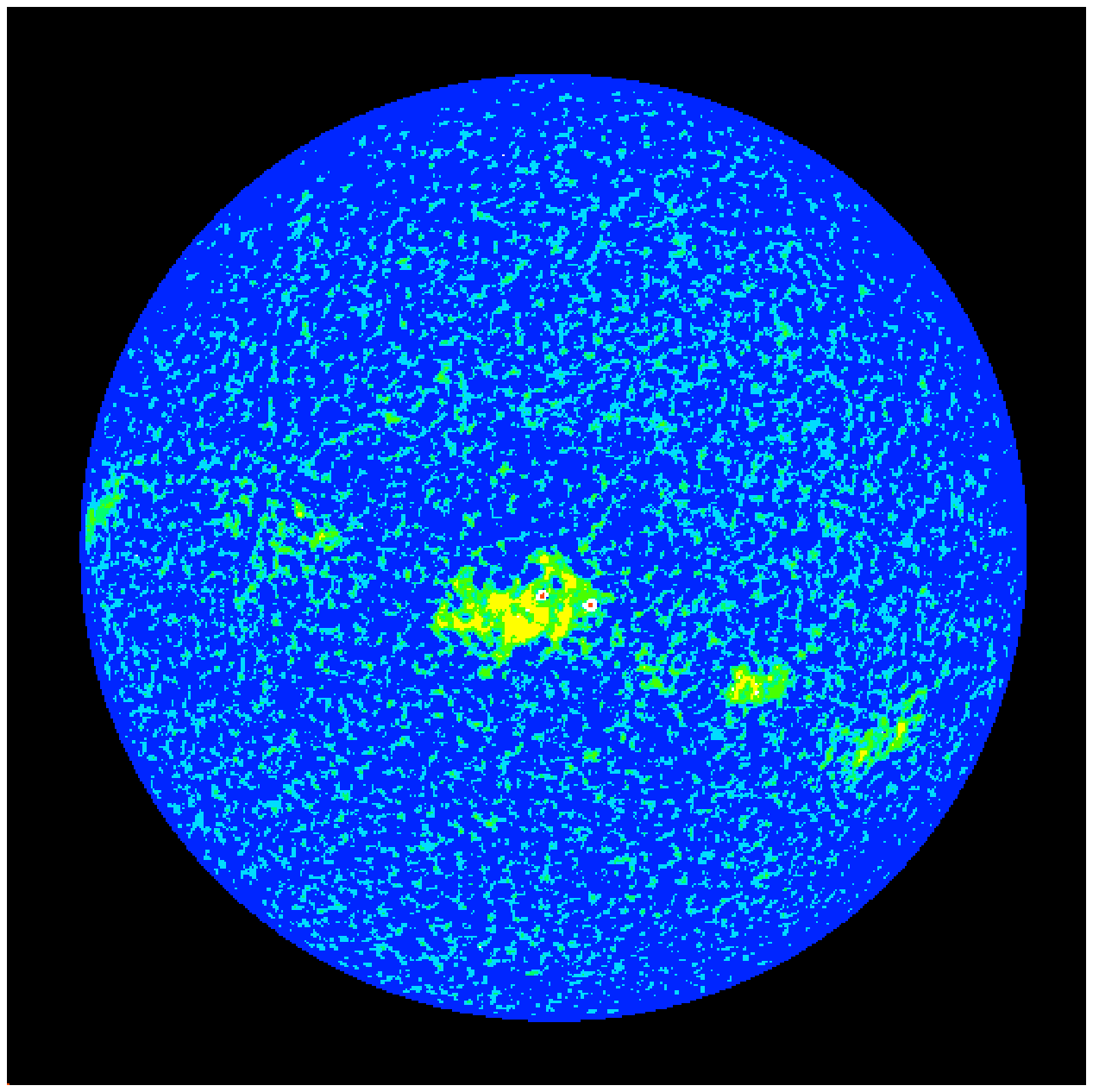}\includegraphics[width=3.5cm]{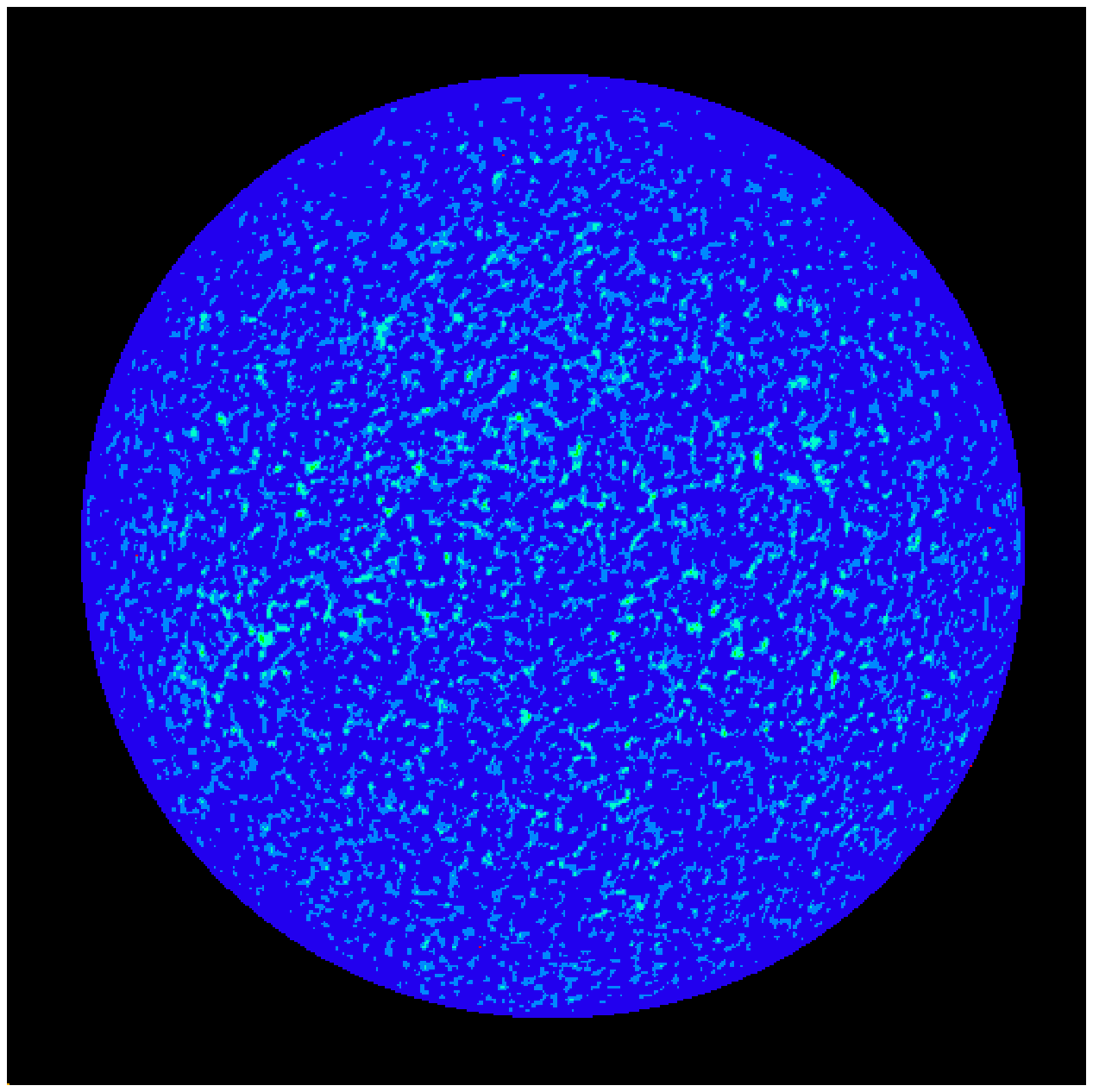}}
\caption{Example of {\it PK-027} (top panels) observations analysed in this study and of the corresponding mask images (bottom panels). 
The observations were taken on 2007 June 7 (left side) and June 21 (right side). The disk features on mask images 
are identified by  color: blue, light blue, green,  and yellow show quiet Sun, network, enhanced network and plage regions, respectively; white 
and red show penumbral and umbral regions (visible only when enlarging the figure).}
\label{fig5} 
\end{figure}

Figure \ref{fig5} shows examples of features identified in two sets of observations analysed in this 
study. 
We measured the radiative emission of the features identified in each image. In particular, we considered the emergent intensity of quiet Sun regions and the contrast 
of bright features. The latter quantity  
is defined, for each heliocentric angle,  
as the ratio $(I_f-I_B)/I_B$, where $I_f$ is the median of intensity values 
of  pixels labeled as feature $f$, and $I_B$ is 
the median intensity of pixels coded as feature $B$ (QS, internetwork).  
The contrast of features was computed in 50 equal-area annuli centered on the solar disk center.  
For each bandpass and feature, the center-to-limb variation (CLV, hereafter) of contrast was computed by taking into account 
the median  contrast values measured for the given feature  in the 50 annuli of all the images in the specified bandpass. 
We note that  we decomposed 
{\it PK-027} images, 
and subsequently measured the radiative properties of identified features  in {\it PK-027},  
 {\it PKR-011}, and {\it PK-010} co-aligned images using this decomposition. 
 This was done to compare  measurement and synthesis results assuming that   the size of 
 identified features is constant  across the various images, i.e.,  
over the atmospheric heights sampled by the {\it PK} filters.

\section{Spectral synthesis}
 
To discuss the results of our measurements and place them in context with results of previous studies, 
we  
computed synthetic spectra in the range of the Ca~II~K line with 
the RH radiative code 
\citep[][]{uitenbroek2002}. This code performs NLTE radiative transfer modeling by utilizing the 
partial frequency redistribution formalism (PRD). 
It
 has been 
utilized for spectral synthesis  and atmosphere diagnostics at chromospheric heights 
 employing various atmosphere models. For example,  
 \citet[][]{tritschler2007}, \citet{reardon2009}, and \citet{pietarila2009} utilized RH and earlier models by \citet{fontenla1993}, while
  \citet[][]{rezaei2008} and \citet{grigoryeva2009} employed the sets of  \citet{fontenla2006} and \citet{fontenla2007}.

Our calculations with RH were performed by adopting a 5-level plus continuum atomic model \citep[][]{uitenbroek1989,uitenbroek2001}, the atomic data of 
\citet{shine1974} with collisional strength values from  
\citet{melendez2007},   
and
the set of 
atmosphere models   
 presented by \citet{fontenla2009}, hereafter FC09, as well as the  earlier models of \citet{vernazza1981} and 
 \citet{fontenla1993,fontenla2006}, hereafter  VAL3, FAL3, and FA06, respectively. 
Since  some of these models were constructed with radiative transfer codes that utilize the 
complete redistribution approximation  (CRD), which is also an option in RH, we 
computed the Ca~II~K  line profiles with both formalisms for comparison. 
This was done to evaluate the sensitivity of synthesis results to code approximations.

 For each reference model, we calculated the emergent spectrum at 43 wavelengths over the range 392.515 nm to 394.214 nm 
  for 
 13 different 
 viewing angles. 
 For each angle and atmosphere, we weighted the obtained  spectrum with the various filter profiles considered in our study and summed up 
 the result over the spectral range of our synthesis. 
The intensities derived with the code were then compared with those measured in {\it PK} images for various disk features. 
The results obtained are presented in  Sect.~4. 

As discussed by \citet{shine1975} and \citet{uitenbroek2002}, we 
found that the Ca~II~K line profiles derived from our simulations using the 
various reference models and adopting either 
 PRD or  CRD,
differ 
in terms of the line relative minima. The amplitude of the line emission peaks
derived for the various models at disk center,  however, is almost unaffected by the adopted redistribution formalism. This is in 
contrast to the suggestion by 
\citet[][ their Fig. 7]{fontenla2009} that the much  higher emission peaks in the H and K lines they found can be attributed to the lack of inclusion of PRD effects. 
We note that the difference in emission peak amplitude between CRD and PRD increases considerably towards the limb \citep{uitenbroek1989} in any given model, and that this
affects the CLV of line-integrated quantities.  We also found that the difference between the profiles derived from the two approximations    
increases according to the brightness of the model disk feature it represents.  
Comparison of emission peaks obtained with both PRD or CRD also indicates that model P differs from the other models in terms of  the smaller 
value of its micro-turbulent velocity.

\begin{figure}
\centering{\includegraphics[width=8.5cm]{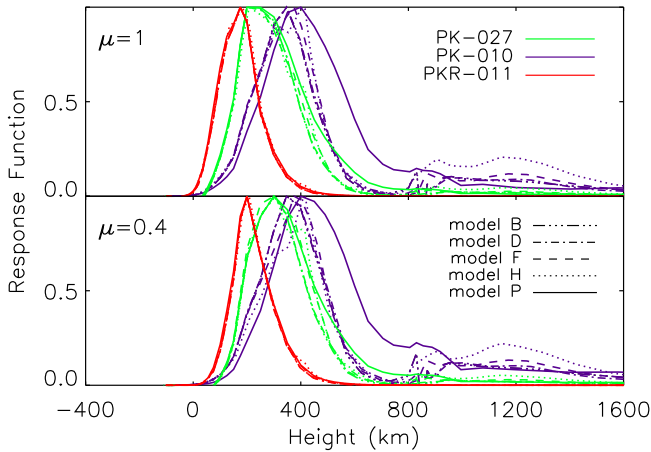}}
\caption{ Intensity response functions to perturbations of temperature for the PSPT  filters computed at the disk positions $\mu=1$ (top) and $\mu=0.4$ (bottom). The functions 
derived for the three filters and  
various atmosphere models are indicated by different colors and lines 
as specified in the legend. Each function is normalized to its maximum. 
} 
\label{fig3} 
\end{figure} 
\begin{table}
\caption{ Characteristics of the {\it PK} filters.}
\footnotesize{
\begin{tabular}{lllll}
\hline
\hline
 & &{\it PK-010} & {\it PK-027} & {\it PKR-011}\\
 \hline
 $\lambda _0$ & (nm) &393.379 & 393.415 & 393.633\\ 
 BW & (nm) &  0.103 &0.273 & 0.106 \\
 \hline
 $\mu$=1 & & & &\\
 Ave  & (km) &375$\pm$25 &200$\pm$5 &  175$\pm$5\\
   FW &(km) &  270$\pm$30&260$\pm$15 &160$\pm$5 \\
RF-f&  & 0.58-0.82 & 0.84-0.94& 0.996-0.998\\
 \hline
  $\mu$=0.4& & & &\\

    Ave & (km) &380$\pm$30 &300$\pm$5 &200$\pm$5 \\
    FW & (km)&270$\pm$35 & 260$\pm$15& 140$\pm$5\\
 RF-f&  & 0.43-0.76 & 0.75-0.91& 0.989-0.995\\
 \hline
\end{tabular}
\tablefoot{For each bandpass analysed,  the table summarizes the central wavelength ($\lambda _0$) and FWHM  (BW) of 
the filter transmission 
profile, the average response height  (ave), and FWHM of response heights (FW) of the various response functions (RFs)
 derived from the atmosphere models of Fontenla et al. (2009) at disk 
positions $\mu$=1  and $\mu$=0.4, and the fraction of RF (RF-f) from atmospheric heights below 500 km at the same disk positions.
} 
}
\label{tab1}
\end{table}	

Next,  
we computed 
the intensity response 
function  (RF) to perturbations of temperature for the {\it PK}. This function (RFs-PK, hereafter) 
allows us to infer   
the atmospheric heights from the intensity 
 measured in {\it PK} images more reliably than other methods \citep{uitenbroek2006}.
 Moreover,  
it 
also indicates the sensitivity of synthesized quantities to atmospheric models,   though 
the CRD and PRD calculations are affected by  uncertainties associated with the static and one-dimensional nature of 
the atmospheres used.
We computed the  RFs-PK only with PRD  because of the advantages that this approach  offers for
 the synthesis of the 
Ca~II~K line \citep{uitenbroek2002} over the  {\it PK} range, where the
Ca~II~K  dominates all the opacity and emissivity sources. 
The filter-response functions of the different atmospheres were calculated numerically as described by \citet{uitenbroek2006}, i.e. by 
perturbing the temperature of an atmosphere at each height, evaluating the emergent intensity integrated over the filter 
bandpass, and comparing the perturbed filter intensities with those of the un-perturbed atmosphere.

Figure \ref{fig3} shows the RFs-PK derived  for the  various atmosphere models at the disk positions   
$\mu=1$ (top) and  $\mu=0.4$ (bottom). 
These RFs-PK indicate that the {\it PK} sample quite a wide range of 
atmospheric heights, and that the radiative signals measured with these filters are dominated by the  wings of 
the 
Ca~II~K line, which form at heights below  500 km. 
We found that   
 {\it PK-010} samples the highest atmospheric regions, while  {\it PKR-011} and {\it PK-027} sample  
middle and upper photospheric heights, respectively. In particular, {\it PKR-011} has the 
sharpest RF and provides  
the cleanest photospheric signal. These properties follow from the filter widths and positions.

All RFs-PK display a smooth variation with atmosphere height, except for  the ones derived 
for model B (representative of QS), which show a sharp variation at a height of $\approx$ 850 km produced by the temperature profile of this model above 
the temperature minimum layers. 
The RF derived for {\it PK-010} and {\it PK-027} also have a 
long tail extending to the middle and upper chromospheric heights 
exceeding  1000 km. However, the response of {\it PK-027} is $\approx$ 3 times lower 
 than that of 
 {\it PK-010} at both heights of 800 km and 1000 km, 
 for  a vertical line of sight and the reference atmosphere model P (bright plage). In addition, while the RFs-PK 
 derived for models B (QS) to model H (plage) exhibit  similar variations with height, 
 the RF-PK computed 
 for model P (bright plage) is  $\approx$ 3 times higher than the RF-PK obtained for model H (plage) 
 at a height of 800 km, but it 
is $\approx$ 3 times 
 lower than the RF for model H (plage) at 1000 km height. 
  
Table  \ref{tab1} summarizes 
the average and 
 FWHM of the response heights of the RFs-PK  for the entire set of atmosphere models.
We found that these quantities 
are only slightly 
sensitive to  the choice of atmosphere models and filter profiles.  
Among the various RFs-PK, the most sensitive to chromospheric conditions are the ones derived for model P and  
{\it PK-010}. 
We also found that the average response height increases from  1\% to 40\% when 
moving toward the limb for {\it PK-027} and {\it PK-010}.

\section{Results}  

\begin{figure} 
\centering{\includegraphics[width=8.5cm]{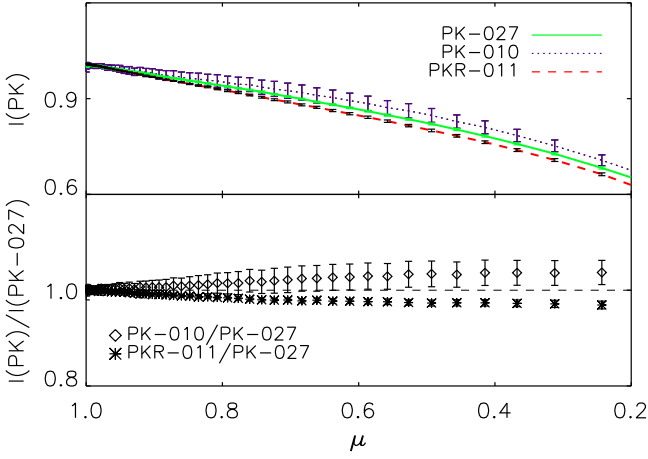}}
\caption{Top: Quiet Sun limb darkening measured in  {\it PK-027} (green,  solid line), 
{\it PK-010} (violet, dotted line),
and {\it PKR-011} (red,  dashed line) good images. 
Intensity values are normalized to the median of values measured at the disk center ($\mu \ge 0.95$).
The error bars show the dispersion of measured values.  
Bottom: Ratio of  the median of quiet Sun limb darkening measured in {\it PK-010} to {\it PK-027} 
(diamond) images, and in  
{\it PKR-011} to {\it PK-027} images (asterisk). 
 The error bars show the dispersion of values from error propagation.  
The dashed line would represent perfect agreement between the two compared sets. }
\label{fig6} 
\end{figure} 
\begin{figure} 
\centering{\includegraphics[width=8.5cm]{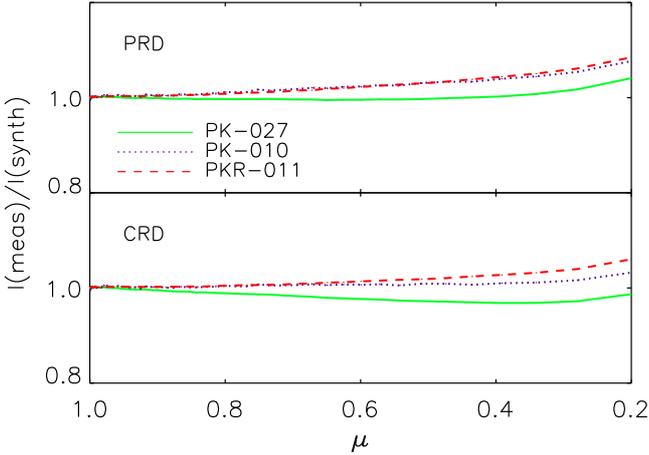}}
\caption{Ratio between the median of quiet Sun limb darkening measured 
in PSPT images and the quiet Sun limb darkening derived from the synthesis performed on model B by FC09 
for the {\it PK-027} (green, solid line), {\it PK-010} (violet, dotted line),  and 
{\it PKR-011} (red, dashed line) filters by assuming either the PRD (top panel) or CRD (bottom panel) formalism.
}
\label{fig7} 
\end{figure} 

Figure \ref{fig6} (top panel) shows the CLV of intensity values  measured for QS
identified in {\it PK}  images. 
In particular, the CLV  computed from the {\it PK-010} images is less steep
than the one found in both {\it PKR-011} and {\it PK-027} observations. Moreover, on average, more poorly  resolved images from 
the poor quality set lead to the CLV  
being larger than  
the median CLV measured for  good quality images, although differences are within the deviation of values measured 
on good images. These findings are in agreement with results obtained for  different Ca~II~K  data sets  by \citet{livingston2008} and   
\citet{criscuoli2008}.

The dependence of the intensity CLV on both filter bandwidth and central wavelength is, however,  small. 
We found that the average and the standard deviation of the  ratio of the values 
measured  
on {\it PK-010} to  {\it PK-027} images
are 1.013 $\pm$ 0.013, for all the disk positions with $\mu \ge$ 0.2 (Fig. \ref{fig6}, bottom panel).
The same quantities derived for the ratio of  the values 
measured  on {\it PKR-011} to {\it PK-027} images 
are 0.99 $\pm$ 0.01. We note that these averages are strongly weighted by the values obtained at positions   
close to disk center, because of the larger number of pixels corresponding to large $\mu$. 

Comparing
the median CLV measured in {\it PK} images 
with the CLV derived from  line
synthesis, 
 we find that the measured CLVs are generally consistent with the results of the spectral synthesis performed on 
the reference atmosphere (model B) utilized in this study. 
In particular, Fig.  \ref{fig7} (top panel)
shows 
this comparison for results derived with PRD.  The differences between measured and modeled values 
increase towards the limb,  
reaching between 4\% and  8\% depending on the bandpass, with the measured limb values always brighter than model predictions.
The average and the standard
deviation 
of the ratio 
of the
measured to modeled values for all the disk positions with $\mu \ge$0.2 are 
0.999 $\pm$ 0.003, 1.015 $\pm$0.015, and 1.013$\pm$0.016 
for the {\it PK-027}, {\it PK-010}, and {\it PKR-011} images, 
respectively. 
The same quantities evaluated for  poor quality images are within the 
deviation of values measured in high quality data. 
These results indicate that the intensity thresholds utilized in this study for the identification of QS
are in fairly good  agreement with those  
obtained with PRD in the reference QS atmosphere  (model B) of FC09.  
Figure  \ref{fig7} (bottom panel)
also shows that the  agreement between measurements and synthesis results slightly  improves by taking into account    
the results from CRD computations, in line with the increase in emission of line emission peaks towards the limb for profiles calculated with
this formalism. We found that the maximum difference between measured and modeled 
values is $\simeq$ 5\%. 
The average and standard deviation of the ratio of  measured to  modeled values obtained with CRD  
over all disk positions with $\mu \ge$0.2 are 
 0.99 $\pm$ 0.01, 1.006 $\pm$0.005, and 1.01$\pm$0.01 
for the {\it PK-027}, {\it PK-010}, and {\it PKR-011}  images, respectively. We discuss  these results in Sect. 5.

\begin{figure}[ht!] 
\centering{\includegraphics[width=8.5cm]{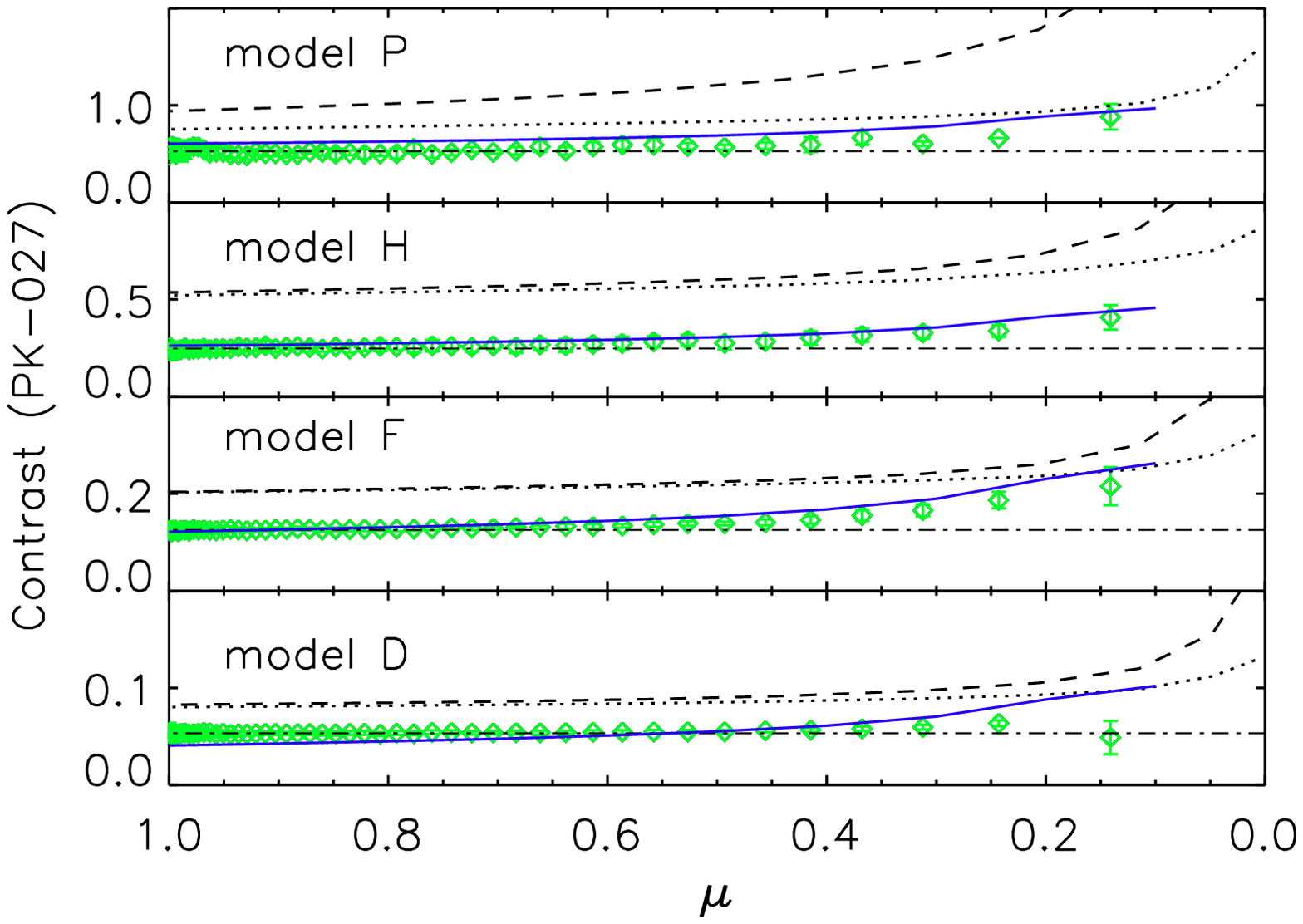}
\includegraphics[width=8.5cm]{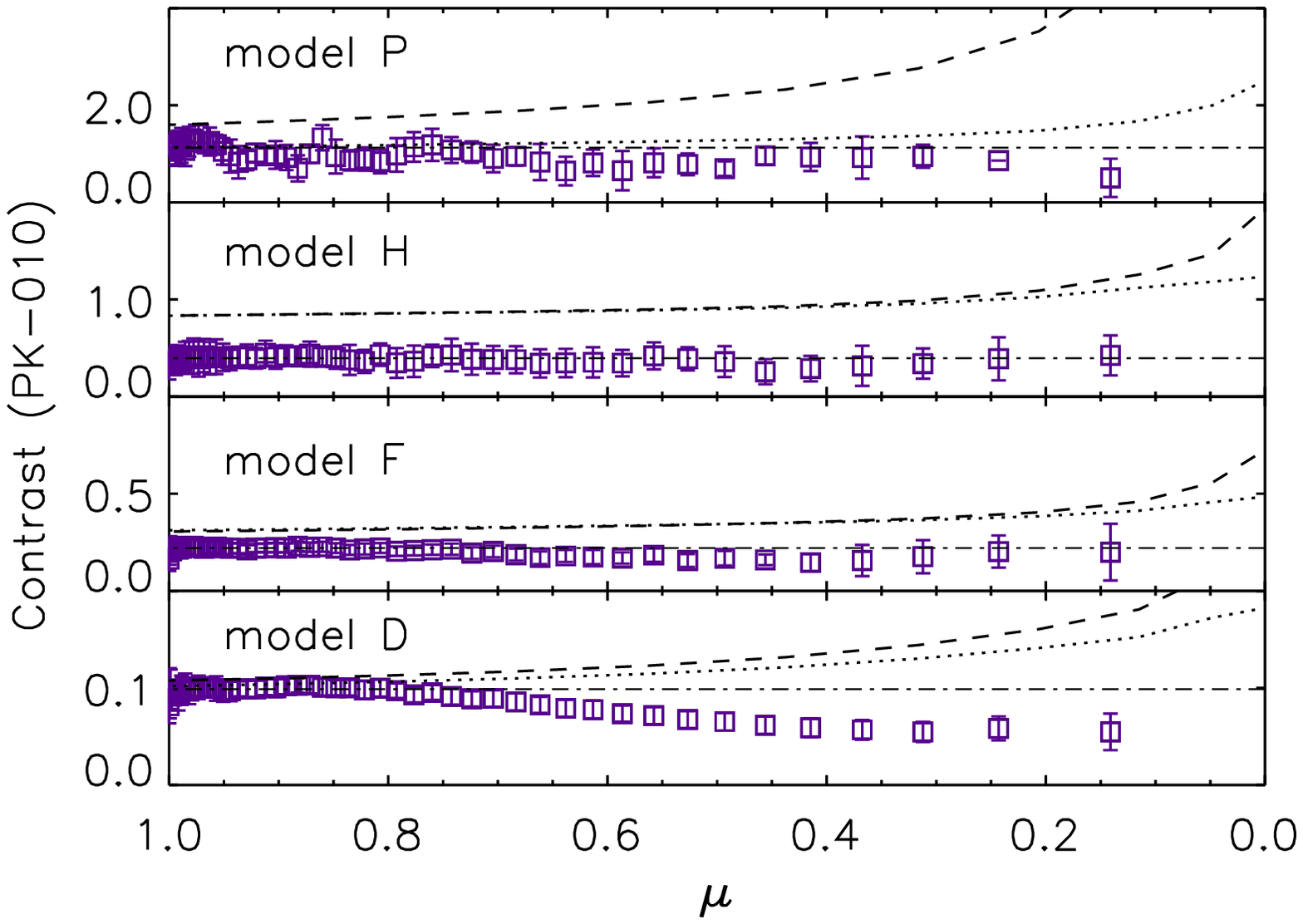}
\includegraphics[width=8.5cm]{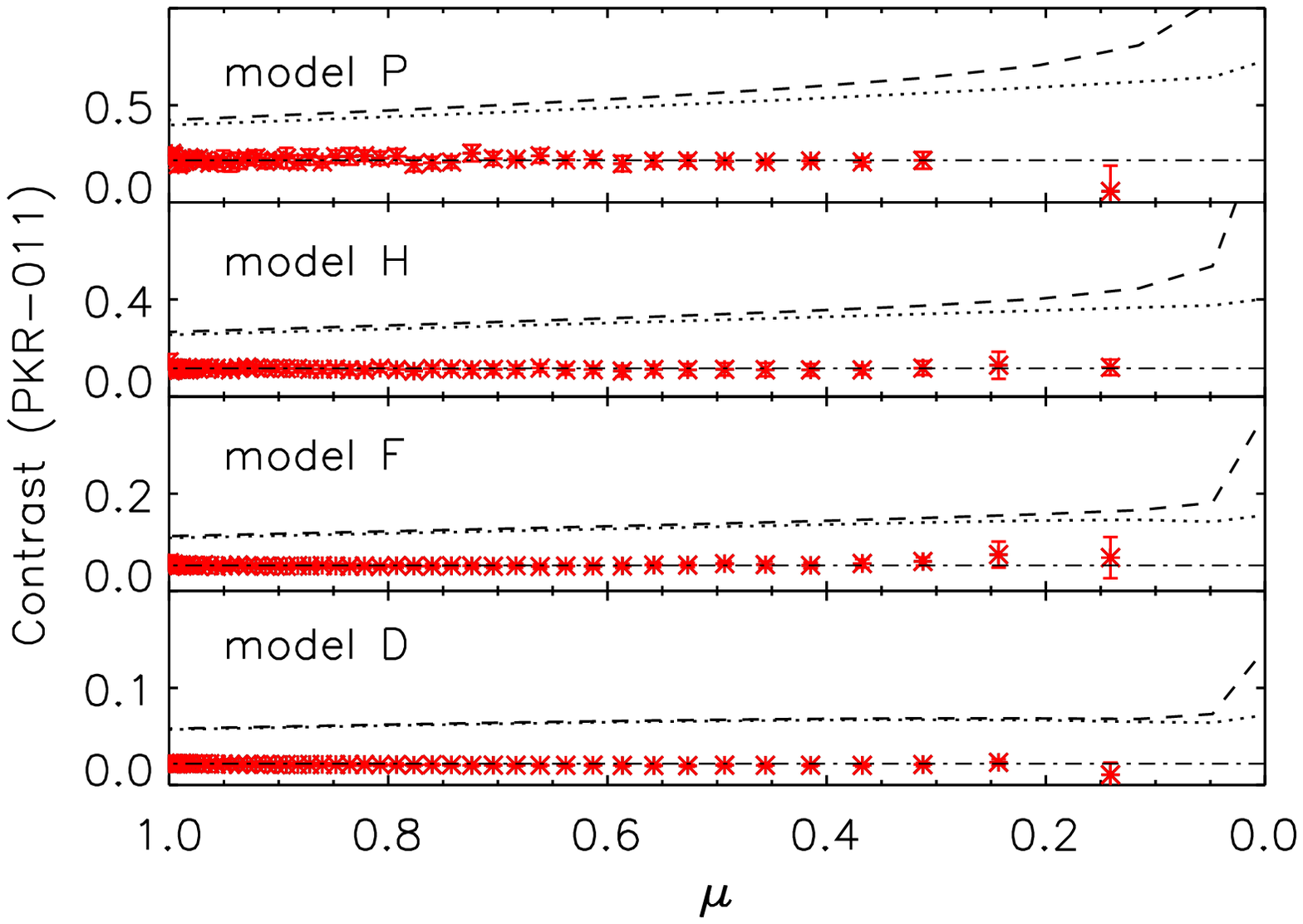}}
\caption{CLV of median contrast values measured  (symbols) for various disk features identified in {\it PK-027} (top panel), {\it PK-010} (middle panel),   
and {\it PKR-011} (bottom panel) images.  For each bandpass, the various sub-panels show 
measurement results for solar features ordered by decreasing contrast, i.e. from the top panel results for  
bright plage, plage, enhanced network, and network regions, respectively. The error bars represent the standard deviation
of measurements. For each disk feature and bandpass, the 
dot-dashed line indicates 
the average of values measured at disk 
positions $\mu \ge$0.9, while dotted and dashed lines show the respective CLV derived from RH calculations with PRD and CRD and 
the reference model corresponding to the given feature, also indicated in the legend. 
In the top panel, the solid blue lines show the threshold values applied to the feature identification.}
\label{fig9} 
\end{figure} 

\begin{figure}[ht!] 
\centering{\includegraphics[width=8.5cm]{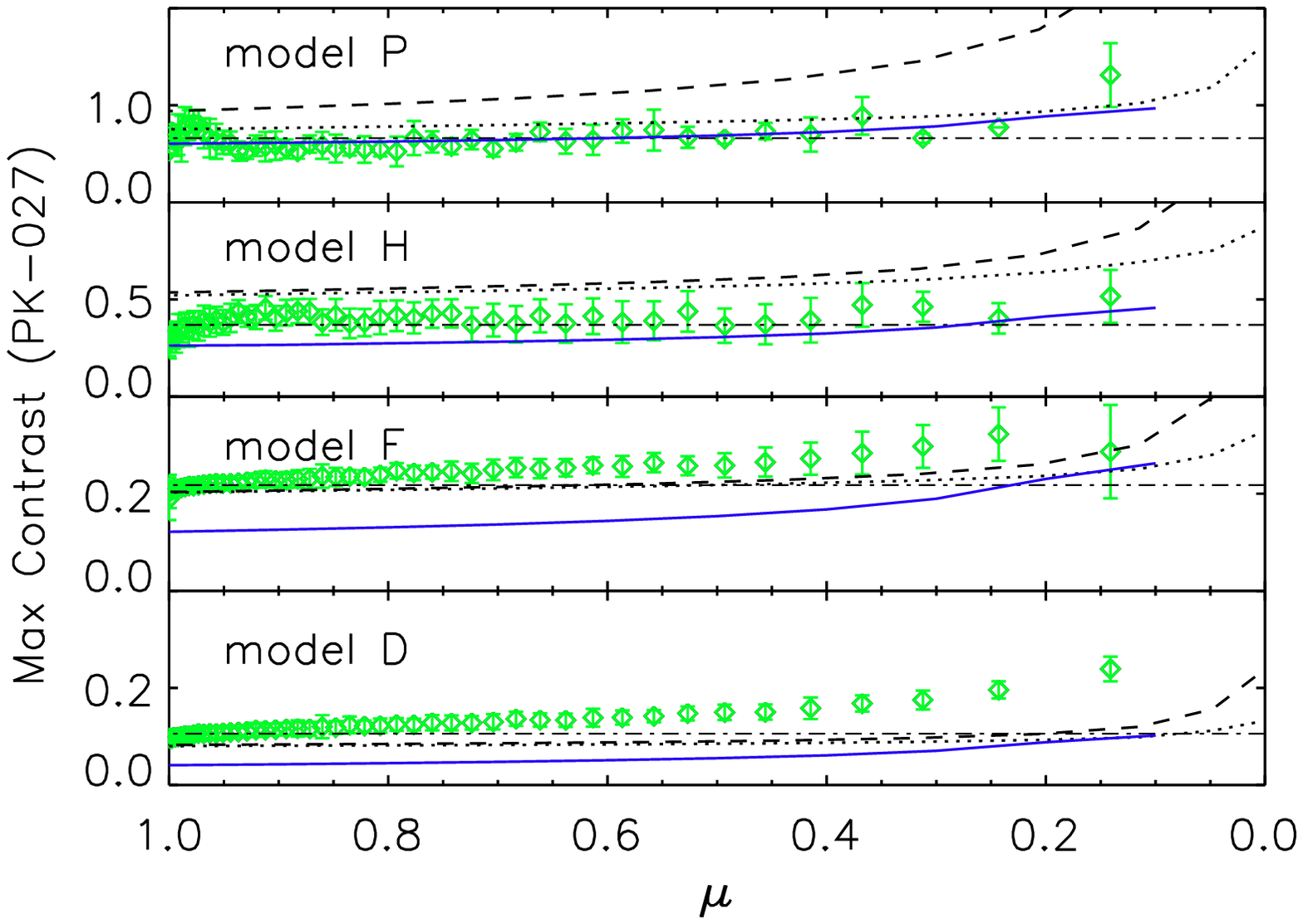}
\includegraphics[width=8.5cm]{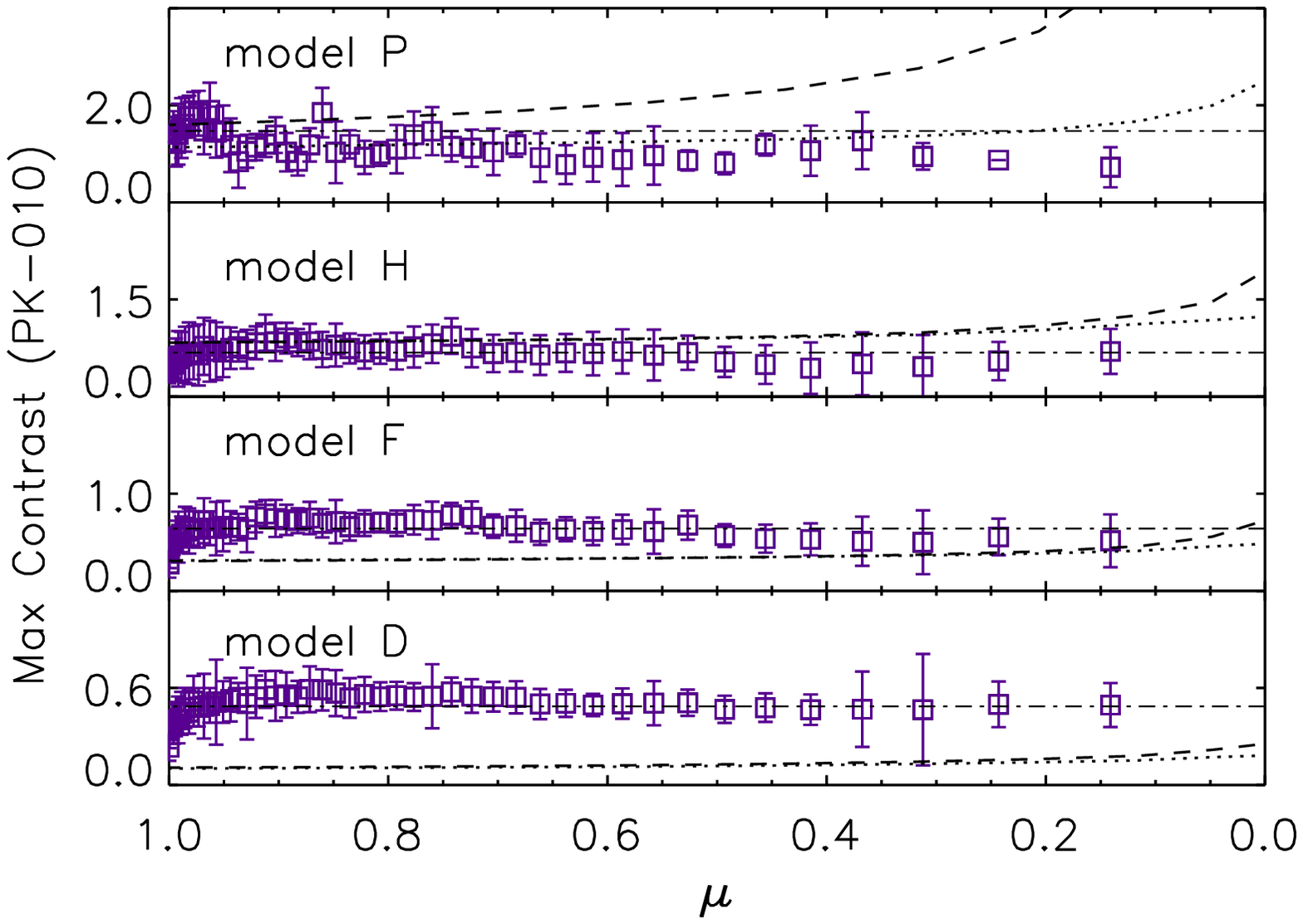}
\includegraphics[width=8.5cm]{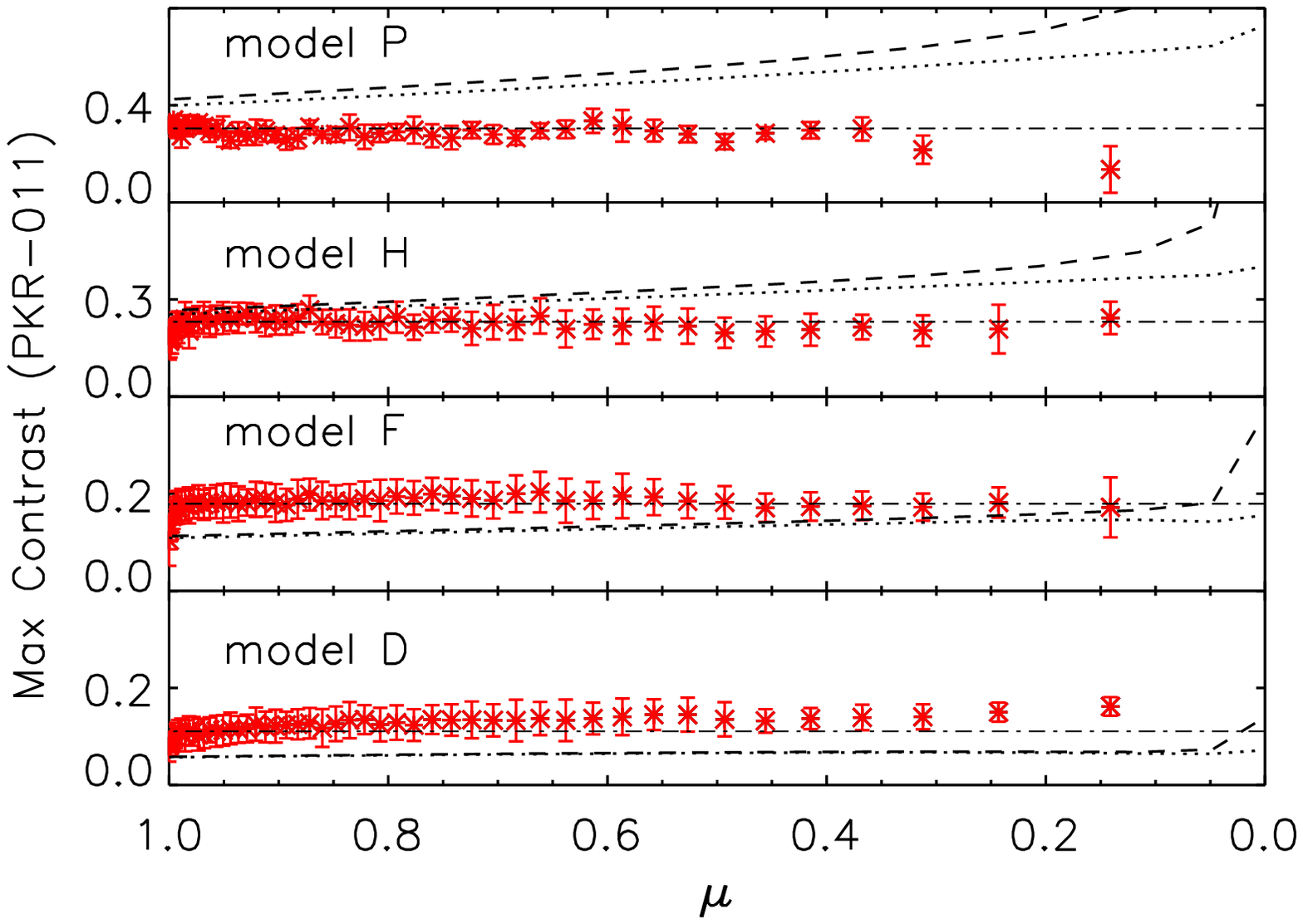}}
\caption{CLV of maximum contrast values measured (symbols) for various disk features identified  in {\it PK-027} (top panel), {\it PK-010} 
(middle panel), and {\it PKR-011} (bottom panel) images. The error bars represent the standard deviation 
of measurements. Legend as in Fig. \ref{fig9}.
}
\label{fig9b} 
\end{figure} 
We now consider the contrast measured for the various bright features identified in the {\it PK} images.  
Figure \ref{fig9} shows that the measured values depend on the type of solar feature, disk position, and filter 
bandpass. In particular, 
contrast values measured for each feature at all disk positions 
in both {\it PK-010} and {\it PK-027} images are higher than those derived from 
{\it PKR-011} observations. This fits in with the well-known observational evidence that the contrast of bright features is greatest at the line core.
However, 
 the contrast values 
measured in the {\it PK-010} images, for the disk positions with $\mu \leq 0.5$, overlap those obtained for the {\it PK-027} observations, which sample  
more extensively the Ca~II~K  
line wings. 

We found that the  median of the measured contrast values, for a given feature over each $\mu$ range, increases towards the limb 
for all the features identified in the {\it PK-027} images 
(by 3\% for the network, by up to 24\% for the plage). 
The same quantities evaluated for all the features identified in
 {\it PKR-011} observations appear to be constant 
across the disk, within the scatter of measurements. This is also true  
 for the results obtained in the {\it PK-010}  images for plage and bright plage features 
(associated with 
 reference models P and H,  respectively).  
On the other hand,   
the network and enhanced network  contrast  measured in the same images
(associated with reference 
models D and F, respectively) show a clear 
monotonic decrease toward the limb. This decrease 
 is likely due to 
 both the limited spatial resolution of analysed images and a foreshortening near the limb. 
We find that the maximum value of contrasts measured for the same features 
and bandpass over each $\mu$ range is constant across the disk, within the scatter of the measurement uncertainty. 
Figure \ref{fig9b} shows the  CLVs of maximum contrast values measured  for the various disk features and bandpasses.
These results  agrees qualitatively with published CLV measurements of bright features from full-disk continuum observations 
\citep[][ and references therein]{ermolli2007,criscuoli2008}.

Figures \ref{fig9} and \ref{fig9b} also compare the measured and synthetic  contrasts  
of models D, F, H, and P.
Figure \ref{fig9} (all panels) shows that  
 the median of the contrast values measured at the disk center are, 
 on average,   
 a factor $\approx$ 1.7
lower than those derived from  PRD using the corresponding
atmospheric models; the discrepancy increases for 
plage regions (associated with model H) identified in all the images and for all the features in {\it PKR-011}. 
Figure \ref{fig9} (top and bottom panels) also indicates that the median  contrast 
values deduced 
from {\it PK-027} and {\it PKR-011} images are in general poorer agreement 
with model predictions 
than  the {\it PK-010} data.

We also find that the values obtained with CRD for the models D to H,  which are representative of network to plage regions, 
at disk center, on average,  
are only $<$ 2\% 
higher than those obtained with 
PRD on the same models, but the same quantity evaluated for model P,  associated with bright plages, is $\approx$ 29\%, 20\%, and 6\% higher in 
{\it PK-010}, {\it PK-027}, and {\it PKR-011} images, respectively. 
When moving toward the limb,  
the discrepancy between 
measured and modeled values slightly decreases for PRD computations, while it increases 
for CRD calculations. 
We  find  that the maximum  contrast values measured for most of the solar features, filter bandpasses, and 
disk positions 
agree in 
general more closely  with the outcomes of the spectral synthesis than the median contrast measurements. In particular, Fig.  \ref{fig9b} 
(top and middle panels) shows that the CLV of the 
maximum contrast values measured for plage and bright plage features identified in  {\it PK-010} images is very satisfactorily reproduced
 by the CLV of contrast values obtained with   
 PRD for the corresponding reference models H and P, respectively. This is also true for   
the same features identified in {\it PK-027} images, and for plages in {\it PKR-011} observations, 
 but with lower fidelity.

\section{Discussion}
\begin{figure}
\centering{\includegraphics[width=8.5cm]{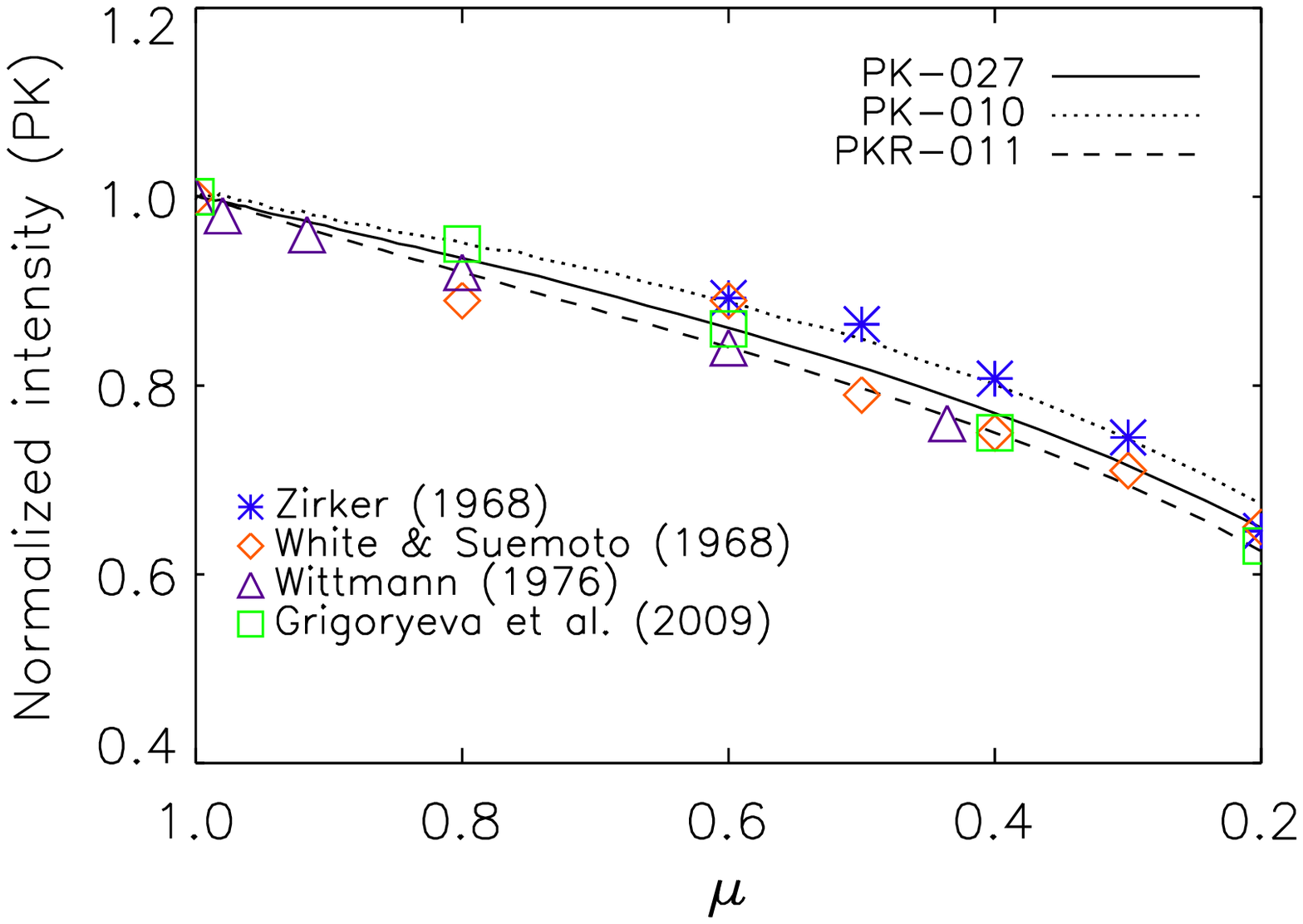}}
\caption{Comparison between intensity CLV of quiet Sun  regions identified in  {\it PK} images (lines, as described in the legend) 
and  published in the literature (symbols, as described in the legend) from observations taken with spectrographs at 
 the Ca~II~K line core. Details are given in Sect.~5.}
 \label{fig13} 
\end{figure} 

The PSPT observations analysed in this study are characterized by high  
 photometric precision \citep{rast2008,goldbaum2009}  and a nearly simultaneous sampling of the solar atmosphere in various 
spectral bands. However, they have only
moderate spatial resolution and  miss the small-scale chromospheric structure found by  
high-resolution observations   \citep[e.g.][]{woger_2006,rezaei2008,cauzzi_2008,scharmer2008}. Thus, PSPT images are ideal for studying 
large-scale phenomena in the solar atmosphere, as in the present study, in which temporally and spatially averaged intensity measurements 
 are compared with  
atmosphere models,   but are less suited to study of the local dynamics, which tend to be averaged at these  pixel resolutions. 
A potential weakness of our study is that 
{\it PK} profiles 
are up to about ten times wider than the bandpasses typically used for Ca~II~K  observations, e.g. of  
Lyot-type filters and spectrographs \citep[e.g][]{kentisher_2008,ermolli2009}, making quantitative comparison with results from narrower band 
observations difficult. 
On the other hand, we note that the bandpass of the {\it PK-010} images is similar to that of the Ca~II~K high-resolution observations 
from the 
Swedish Solar Telescope \citep{scharmer2003} analysed e.g.
by \citet{pietarila2009}, while the bandwidth of {\it PK-027} data is close to that of the  Ca~II~H high-resolution 
images, from the SOT/BFI instrument \citep{tsuneta2008}  onboard the Hinode spacecraft, studied e.g. by \citet{lawrence2010}.

To assess the importance of the filter width, we compare our measurements with results available in the literature 
for observations of the QS taken 
with spectrographs at 
the Ca~II~K  line core, i.e. close to the spectral range sampled with the {\it PK}, but with a narrower bandwidth than that of {\it PK}. 
Figure \ref{fig13} shows that our 
CLV measurements from {\it PK-010} images are 
in striking agreement 
with the results presented by \citet{zirker_1968}. Moreover, our measurements from {\it PK-027} and {\it PKR-011} images, i.e. those 
obtained at the line core with the widest bandpass  
and in the red line wing,  reproduce
the results published 
by \citet{white_suemoto1968}, \citet{wittmann1976}, and \citet{grigoryeva2009} fairly well. 
 Thus, the intensity  CLV of QS  appears to be relatively  insensitive to the exact 
width or central wavelength of the Ca II K filter. 
However, it also implies that the measurements
by \citet{white_suemoto1968}, \citet{wittmann1976}, and \citet{grigoryeva2009} 
were derived from data obtained in different  
observational 
conditions (e.g. spectral sampling, spatial resolution, seeing) and after different types of processing  
 (e.g. data calibration, straylight compensation) than in \citet{zirker_1968}.  

The close agreement found between our measurements and published results 
indicates that any systematic errors affecting our QS measurements, originating in the data 
analysed or caused by image processing applied, do not exceed the  
scatter in previous measurements. 
 We expect that this also holds  for the measurements of the bright features identified at the disk center. 
However, at the limb, our measurements do not account for the dependence of feature size 
 on both atmosphere height and disk position. 

To investigate the effects of feature misclassification on our measurements, we applied a second independent decomposition 
method to the data, specifically that developed by \citet{nesme1996}. 
A direct comparison of the  results obtained with the two methods is hampered by 
the different number of feature classes. 
However, 
the results obtained from this test support those presented in Sect. 4. 
The contrast values measured for the features identified by the test-processing are, on average,  
lower than those presented in Sect. 4 and in poorer agreement with the
results derived from the synthesis at all disk positions. 

\begin{figure}[t!]
\centering{\includegraphics[width=8.5cm]{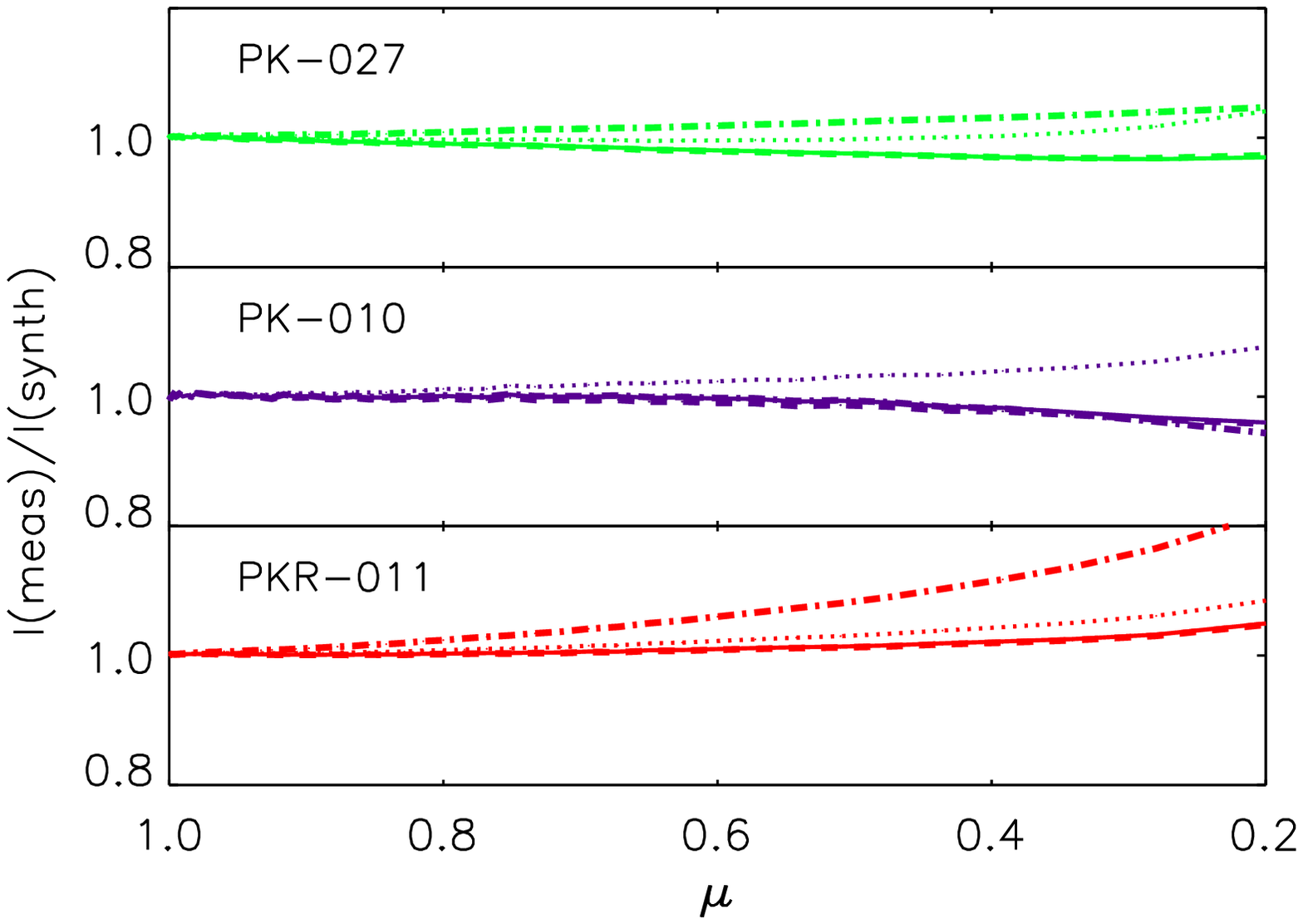}}
\caption{Ratio of the median CLV measured for quiet Sun regions in  {\it PK-027} (top panel), {\it PK-010} 
(middle panel), and {\it PKR-011} (bottom panel) images  to the CLV
derived from the synthesis performed  
with PRD using various atmosphere models. Dotted, solid, dot-dashed, and dashed lines show results derived from the quiet Sun models 
presented by \citet{fontenla2009},
 \citet{fontenla2006},  \citet{vernazza1981}, 
and \citet{fontenla1993}, respectively.}
\label{fig14a} 
\end{figure} 
\begin{figure}[t!]
\centering{\includegraphics[width=8.5cm]{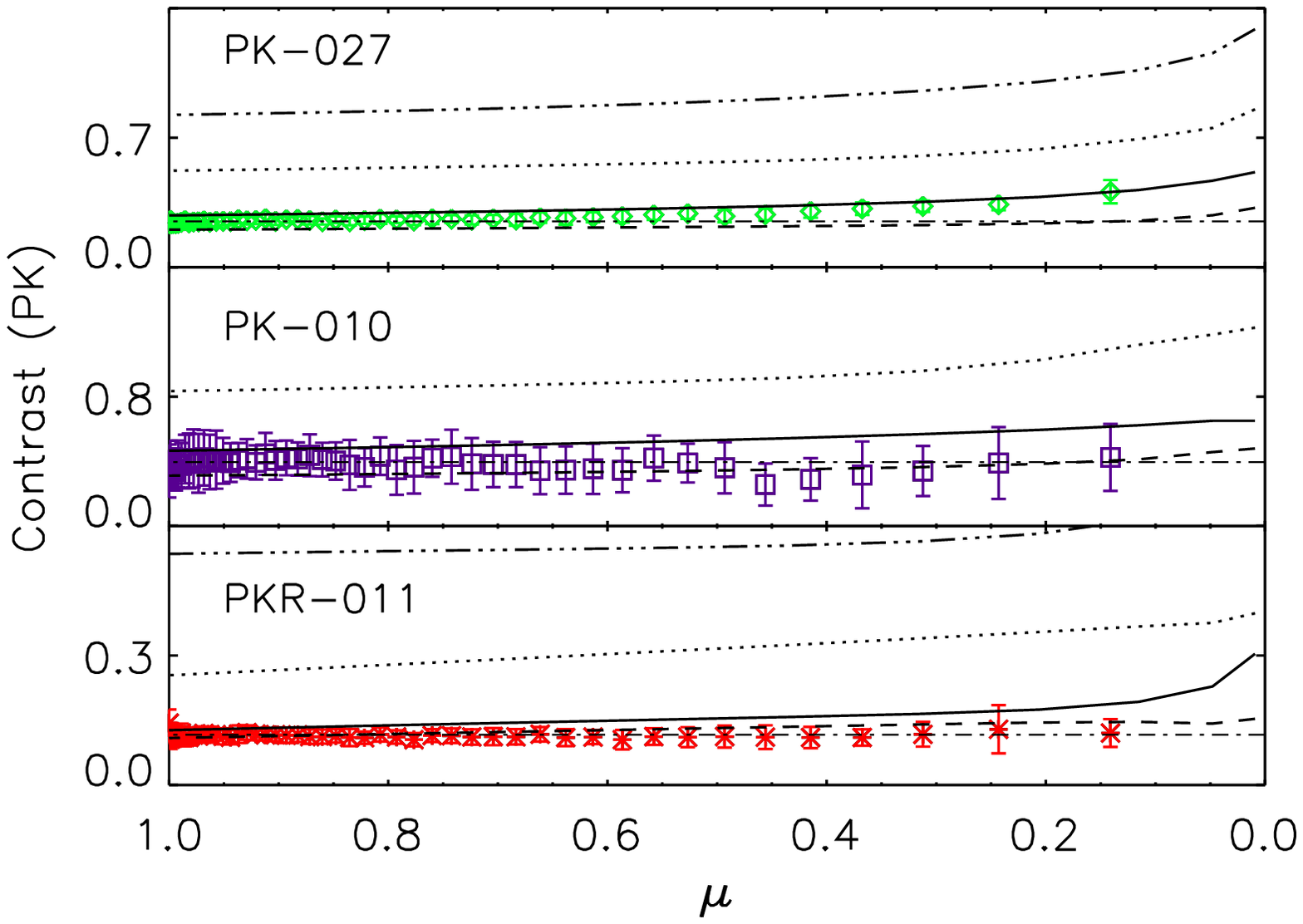}}
\caption{CLV of median contrast values measured   (symbols) for plage regions in {\it PK-027} 
(top panel), {\it PK-010} (middle panel), and {\it PKR-011} (bottom panel) images. For each bandpass, the 
dot-dashed line indicates the average of values measured at disk positions $\mu \ge 0.9$, while dotted, dashed, solid, and triple-dot-dashed 
lines show the respective CLV derived from calculations with PRD for models H and F by 
\citet{fontenla2009}, and for models F and P by \citet{fontenla1993}, respectively.  For 
{\it PK-010}, synthesis results 
for  model P of \citet{fontenla1993} lie outside the 
plotted range.}
\label{figfalf} 
\end{figure} 

In the present study, we
 compared  PSPT measurements with the outcome of spectral synthesis performed with a  well-validated code
on the FC09 atmosphere models. 
 We are aware that none of 
these models can realistically reproduce  the 
very inhomogeneous atmosphere depicted  by high-resolution observations. Some     
spatial and temporal scales of atmosphere patterns displayed by high-resolution observations, as well as some amplitudes of temperature fluctuations, 
can be partly 
reproduced only by hydrodynamic models 
\citep[e.g.][]{carlsson_stein_1995,carlsson_1997,wedemeyer_2004,hansteen2007,leenaarts2009}. 
However, other temporally and spatially averaged properties of the solar atmosphere deduced from observations, 
e.g. the ones discussed by \cite{fontenla2009}, 
  are reproduced fairly 
  precisely 
 by the 
   static and one-dimensional models, such as those considered in our study.  
   As discussed by \citet[][]{rezaei2008}, 
   these models can be 
  employed to derive a number of diagnostics 
   (e.g. response functions to temperature perturbations) 
  and to represent temperature and  density 
  variations  
  as a function of height in the solar atmosphere, at least across the spectral range and 
  at the temporal and spatial resolutions of observations analysed for our study. 

We find that the agreement between measured and modeled results for the intensity CLV of
QS varies with the
redistribution formalism  used in the  spectral synthesis. 
In particular and somewhat surprisingly, the measurements are, on average,  in slightly closer 
agreement with CRD computations than  the PRD calculations.
This may  be because the atmosphere models utilized in this study were computed assuming CRD
 \citep{fontenla2009}. 
If so, CRD, which results in enhanced emission towards the limb, compensates for the steep CLV that 
would result from the low density chromospheric temperature if the more physically realistic formulation of 
PRD were included. 
Therefore, the reference model   
needs to be modified  to fit the data using the PRD formulation.  

To investigate the effects of adopting different reference models on the results obtained,  
we compared measured values for the various disk features and bandpasses with the outcomes of the 
synthesis performed  with PRD on earlier sets of semi-empirical models.  
In particular, we analysed 
the VAL3  and FAL3 
 series, and their subsequent upgrade FA06. 
Figure \ref{fig14a} shows the ratio of  the CLV of median QS intensity values measured in {\it PK} images to the CLV derived from  
PRD computations for the quiet Sun models of the series indicated above. 
For all bandpasses, the intensity measurements are best reproduced by the synthesis performed on
the QS model 
of FAL3 series (FAL3-C). For some bandpasses, 
synthesis results for the QS models of both  
VAL3 and FA06 series (VAL3-C and FA06-C, respectively) 
 reproduce measured values more closely than achieved with the QS atmosphere (model B) of 
FC09. In addition, 
computations  using the bright feature models of the FAL3 set (FAL3-F  and FAL3-P)
reproduce the contrast measurements   
for plage and bright plage features (associated with models H and P of FC09, respectively) in the {\it PK-010} images,
 within the standard deviation of measured values.
In particular, Fig. \ref{figfalf} shows that computations using the FAL3-F model
reproduce the contrast measured for plage features  
in all the {\it PK} images, within the standard deviation of measured values  over many of 
 the disk positions analysed. Figure \ref{figfalf} also shows that the same measurements are 
 reproduced by the computations using model F of 
 FC09  more closely than those derived from model H of the same set,  which implies that the correspondence between 
 models and associated features could be improved yet further. 
 
These findings both confirm and extend the results  presented by \citet{grigoryeva2009}. These authors 
compared measurements from spectroscopic observations of QS taken in the spectral range containing the 
Ca~II~H and K lines, 
with computations performed with 
RH and PRD using various atmosphere models. In particular, they analysed the 
VAL3-C and FAL3-C models, as well as the QS model presented by \citet{fontenla2007}. 
\citet{grigoryeva2009} found that this latter model, 
which is quite similar to model B  of the FC09 set utilized here, 
most closely reproduces their intensity CLV measurements in the Ca~II~K  line wings, while at the line core the synthesis results for this model 
are in poorer agreement 
with measurements than results for earlier models. We note, however, that our study differs from that carried out by   \citet{grigoryeva2009}
in several aspects. In particular, we analysed observations of various solar features, mostly observed in the Ca~II~K  line core, 
while 
the other authors restricted their study to QS mainly observed in the line wings. 
Moreover, we  analysed the most recent set of atmosphere models presented in the literature, in addition to 
the earlier series, while only the latter 
were  
considered by \citet{grigoryeva2009}. Finally, we utilized both the PRD and CRD approximations of the spectral synthesis. 
This allowed us to investigate the effects of code approximations and adopted models. 

Discussing the results obtained, we also consider that the {\it PK} images are contaminated by low levels 
of straylight. By checking the falloff of intensity just outside the solar limb, 
we found that the  images analysed in this study show the same contamination level as that estimated by \citet{criscuoli2008} 
in Ca~II~K  observations from the Rome-PSPT. In particular, we found that the ratio of  azimuthally averaged 
intensity 
values
measured at 1.2 solar-disk radii to the intensity measured at disk center is  0.0167$\pm$0.0005, 0.0150$\pm$0.0005, and 
0.0120$\pm$0.0005  for {\it PK-010}, {\it PK-027}, and 
{\it PKR-011} respectively.  Therefore, following \citet[][]{criscuoli2008} we expect an increase of the order of 30\% in feature contrasts,  
if they are measured in 
{\it PK} images restored for straylight degradation.  

To evaluate the effects of straylight degradation on the comparisons presented above, we developed two test analyses. Firstly,  
we studied straylight degradation effects on synthetic solar images. In particular, 
we applied a model point 
spread function (PSF) to synthetic images 
 derived from our calculations with RH and the FC09 models. We measured the decrease in the intensity and contrast values derived from these 
 synthetic images, hereafter synthetic atmospheres.  This provided a rough indication of the influence of straylight contamination on 
measured contrast values, by avoiding photometric uncertainties introduced by the iterative process of image-restoring 
methods. 
We also  
repeated our measurements for all the features considered in this study in {\it PK} images restored from straylight degradation.

The details of the test analyses we performed are given in the appendix.  To summarize, we found that 
the
 convolution of the model PSF with homogeneous atmospheres, i.e., synthetic atmospheres composed 
 of a single feature at all disk positions, only
leads to a significant decrease in intensity  and contrast values modeled for the brightest 
features  close to the solar limb. 
On the other hand,  convolution of  inhomogeneous atmospheres  with the same PSF, i.e., synthetic 
atmospheres 
composed of quiet Sun and a number 
of bright features at various $\mu$ positions, leads to a significant decrease in 
 contrast values estimated for all the features and disk positions. In particular,  
 convolution of the PSF with synthetic atmospheres 
 composed of 10 small features, each covering 0.02 solar-disk radii, on average, leads to a decrease by 
 nearly a factor two in the modeled contrast values. This value is similar to the 
 mean 
 discrepancy  
 between measured and modeled results for the various bright features 
 presented in Sect. 4. 
In agreement,  
 image restoration from straylight degradation leads 
to an average increase of $\approx$ 18\% over  
all the solar features and a slightly closer agreement between measured and modeled values. 
Model predictions and contrast values measured in restored images at disk positions $\mu \ge$ 0.2 differ by a factor ranging  
from 0.9 to 2.7 (mean  is $\approx 1.6$), depending on the solar 
feature and filter bandpass, this compared to a factor of from 1.1 to 3.1 (mean of $\approx$ 1.9) for un-restored images.

While the results of our test analyses indicate that modeled and measured values are in slightly closer agreement after 
 taking into account the effects of straylight 
contamination, the model predictions  
are still approximately a factor   1.45  higher than measurements obtained for all the solar features at disk center in restored images.


\section{Conclusions}
We have presented a study of the radiative emission of solar features across the Ca~II~K spectral 
range.  We have analysed moderate-resolution PSPT observations taken with interference filters that sample the Ca~II~K range with different 
bandpasses. 
We have compared the results of CLV measurements for various disk features  
with the outputs of spectral synthesis performed with a well-validated radiative code and the  most recent set of 
semi-empirical atmosphere models presented in the literature, as well as with  earlier similar models. 
This led to the following results: 

(1) In general agreement with results presented in the literature, we found that 
the CLV of intensity values measured in the Ca~II~K  range for quiet Sun regions depends on both the filter bandpass and image quality. However, the measured dependence is quite small. 
The quiet Sun's CLV obtained in this study from analysis of  the
{\it PK} images 
is in striking agreement with published results from observations taken with spectrographs.

(2)  The intensity response function for temperature perturbations derived for all the filters considered in this study depends
 only slightly on the filter bandpass and 
reference atmosphere; the most sensitive to higher atmospheric layers are the ones derived for the filters with the narrower-band and the 
reference atmosphere of a bright plage (model P).
This model, however, differs from the other models adopted in this study, not only in terms of its higher chromospheric temperature, but also 
its smaller value of  micro-turbulent velocity. We note that even for the narrowest filter, {\it PK-010}, 
 58\%-82\% of 
 the contribution at disk center 
 is from atmosphere heights below 500 km, with the fraction
being lower for brighter features, while the same quantity evaluated for {\it PK-027}, which has a 
 bandwidth similar 
 to that of the SOT/BFI Ca~II~H  filter onboard Hinode, is 
84\%-94\%. 

(3) RH calculations with the CRD approximation of the quiet Sun atmosphere (model B) 
of \citet{fontenla2009}, on average, reproduce 
measurement results of intensity CLV for quiet Sun regions slightly more closely  than 
PRD computations. 
However, 
RH calculations with PRD for   
earlier models of the quiet Sun atmosphere reproduce measured quantities with a similar accuracy.

(4) 
The 
median contrast values measured for most identified features, bandpasses, and disk positions are, on 
average,  
a factor $\approx$ 2 lower than the contrast values derived from RH for
the reference models considered in this study. Computations in PRD display a closer agreement with the 
data, which, however, 
is still unsatisfactory.
Calculations for older sets of bright feature models 
 \citep{vernazza1981,fontenla1993,fontenla2006}  
fit  measurements from plage and bright plage regions as well as those derived 
from the corresponding new atmosphere models. 

(5) For most of the identified features, disk positions, and filter bandpasses analysed in this study, the 
 maximum contrast values measured in {\it PK} images are in general closer  agreement with model predictions
 than for median contrast measurements.

(6) 
The mean discrepancy between modeled and measured values reduces, on average, by about 12\% 
after taking into account straylight degradation effects on PSPT observations.

In conclusion, the success of CRD computations of  the new atmospheres in reproducing some of the quantities analysed in 
this study does not prove conclusively that 
the radiative code used in this approximation or the adopted models realistically reproduce the observed quantities. However, it 
supports the combined use of this formalism and models, as carried out for the
spectral modeling 
of disk features over wide wavelength ranges \citep[e.g.][]{fontenla2009}. 
On the other hand, the synthesis performed with PRD, on average,
 produces radiative estimates that are in closer  agreement with measurements 
 achieved in this study from bright disk features than  those obtained with the less accurate CRD 
 computation, in particular for the observations taken at the line core. This applies equally-well 
 to synthesis results derived using either the most recent or the earlier atmosphere models, thus supporting simulations 
 performed with PRD  on earlier atmosphere sets 
 for diagnostics across  narrow spectral ranges of the Ca~II~K  line \citep[e.g.][]{reardon2009,pietarila2009}. 
The results derived from our investigation also indicate that 
 the neglect of PRD, 
 to some extent, can be compensated  for by a change in the reference
  models. This is very similar to the ambiguous situation reported by \citet{rutten1982} and \citet{shchukina2001} in the modeling of Fe I lines in 
  the photosphere, where  
  LTE and NLTE calculations can equally well fit the line profiles, as long as the temperature gradient in the atmosphere is adjusted separately in
   each case.

\acknowledgements

We are grateful to Juan Fontenla for the helpful and fruitful discussions. We also thank the International Space Science Institute (Bern) 
for giving us the opportunity to discuss this work with
the international team on "Interpretation and modeling of SSI measurements". 
This work was supported by the Istituto Nazionale di Astrofisica (INAF), the Agenzia Spaziale Italiana, grant ASI/ESS I/915/01510710, sub-task 2210, 
and by the the WCU grant No. R31-10016 funded by the
Korean Ministry of Education, Science and Technology.

\appendix
\section{Straylight contamination}
To evaluate the effects of straylight degradation on the comparisons between modeled and measured 
quantities presented in Sect. 4, we developed two test analyses. 
Firstly, we measured the effects of the straylight contamination induced by a model point spread function (PSF) 
on synthetic solar images.  The PSF 
used in our analysis  
follows \citet{criscuoli2008};  it  
includes  one Lorentzian and three Gaussian components, which are defined by  seven parameters. We  fixed six parameters 
of the PSF components derived by \citet{criscuoli2008}, 
while  
setting the remaining parameter to reproduce the average value of the straylight level measured in {\it PK} images at 1.2 solar-disk radii. 
This  parameter
describes the  weight of the Lorentzian relative to the Gaussian components. 

\begin{figure} 
\centering{\includegraphics[width=8.5cm]{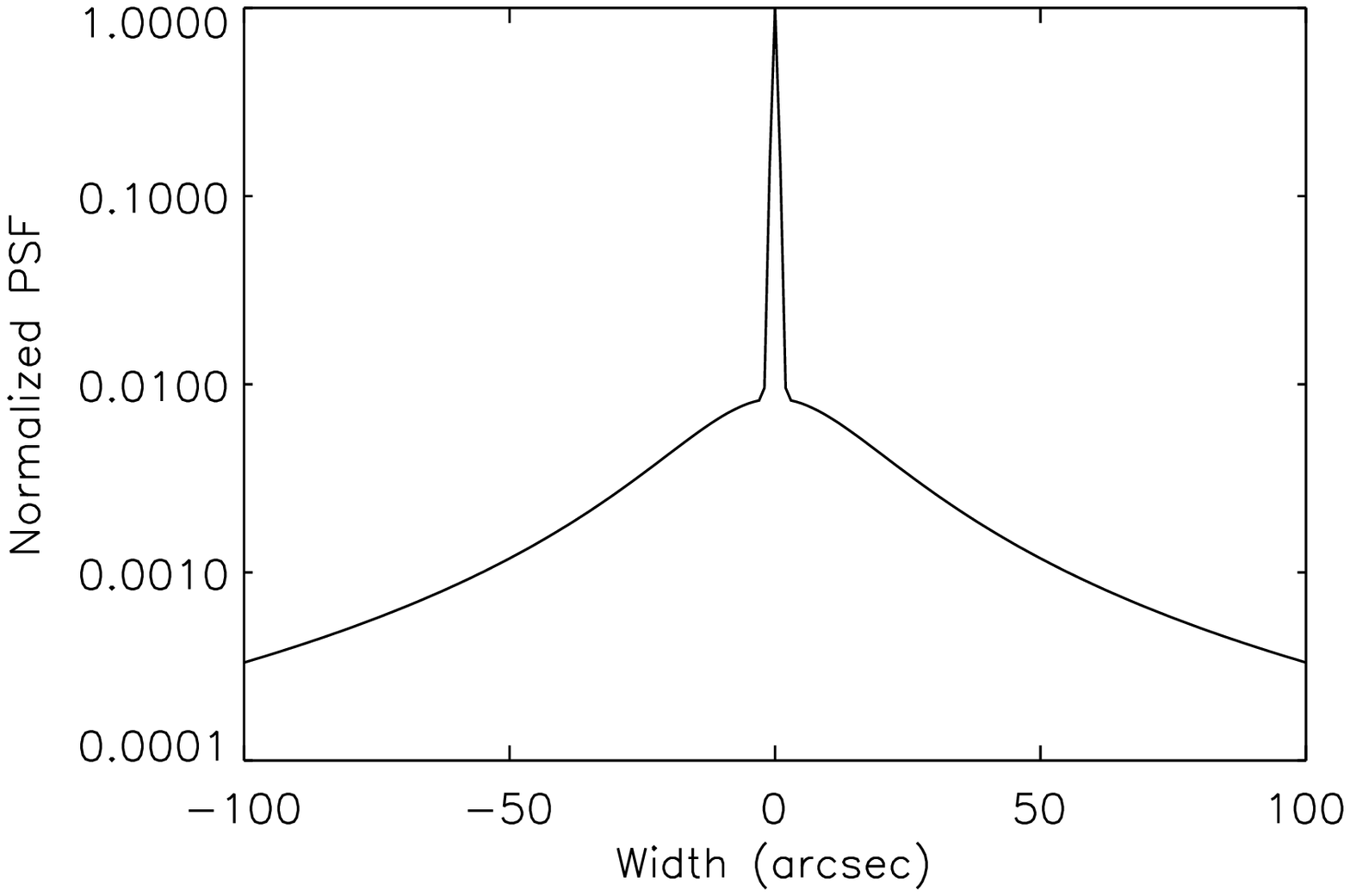}}
\caption{The model PSF used to evaluate effects of straylight degradation on {\it PK} images follows \citet{criscuoli2008}. It
 includes three Gaussians and one Lorentzian component. 
}
\label{reffiga1} 
\end{figure}

\begin{figure} 
\centering{\includegraphics[width=8.5cm]{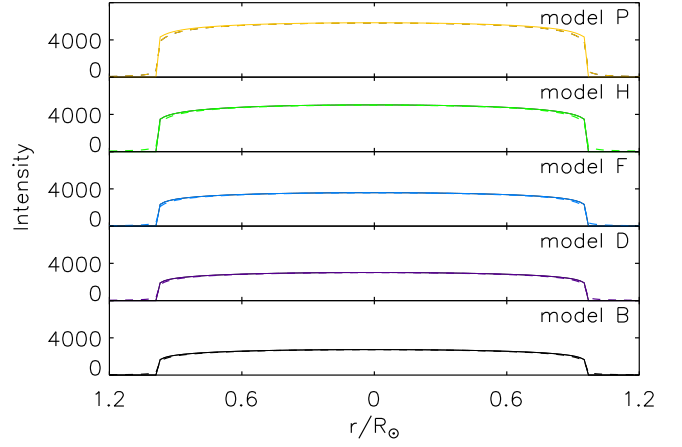}}
\caption{Profiles of intensity values in the set of synthetic homogeneous atmospheres (solid lines) computed for the various 
disk features 
and {\it PK-010} bandpass. 
In the various sub-panels, the different atmospheres are identified by colors and labels: yellow, green, blue, violet, and black
show bright plage, plage, enhanced network, network, and quiet Sun atmospheres, associated with models P to B, respectively. 
Distance from disk center is given in disk radius units. The dashed lines show the profiles of intensity values 
in the various atmospheres after their convolution by the model PSF. 
}
\label{reffiga2} 
\end{figure}

\begin{figure} 
\centering{\includegraphics[width=8.5cm]{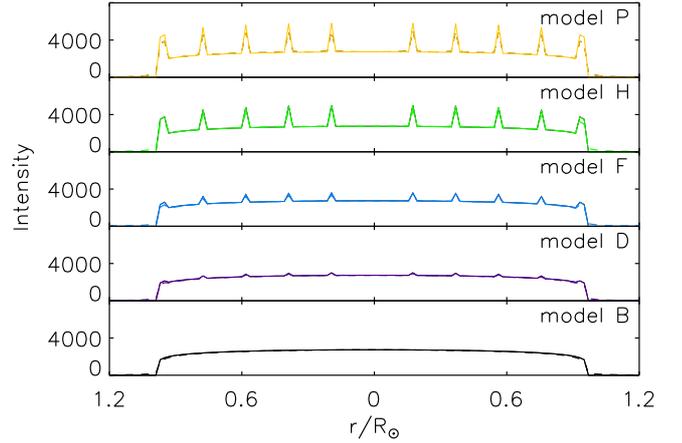}}
\caption{Profiles of intensity values in the set of synthetic inhomogeneous atmospheres (solid lines) computed for the various 
disk features 
and {\it PK-010} bandpass. Legend as given in Fig. \ref{reffiga2}. The dashed lines show the profiles of intensity values 
in the various atmospheres after their convolution by the model PSF.  
Distance from disk center is given in disk radius units. 
}
\label{reffiga3} 
\end{figure}

Figure \ref{reffiga1} shows the model PSF utilized in our analysis.  
We convolved  synthetic images constructed from intensities calculated with the one-dimensional  FC09 models 
for the various disk features and bandpasses analysed in our study with the same model PSF. 
We constructed two sets of synthetic atmospheres. 
The first set is composed of homogeneous atmospheres. 
The second set contains inhomogeneous atmospheres with two components, each atmosphere displaying quiet Sun and a number of features 
of the same type 
at various $\mu$ positions. We describe in the following the results obtained by assuming 10 features, each 
covering 0.02 solar-disk radii at all $\mu$ positions, i.e. foreshortening was not taken into account. 
For the various FC09 models at the {\it PK-010} bandpass, Figs.
 \ref{reffiga2}  and \ref{reffiga3}  show the intensity values computed at various $\mu$ positions in the original synthetic 
 atmospheres and  the atmospheres convolved by the model PSF.

\begin{figure} 
\centering{\includegraphics[width=8.5cm]{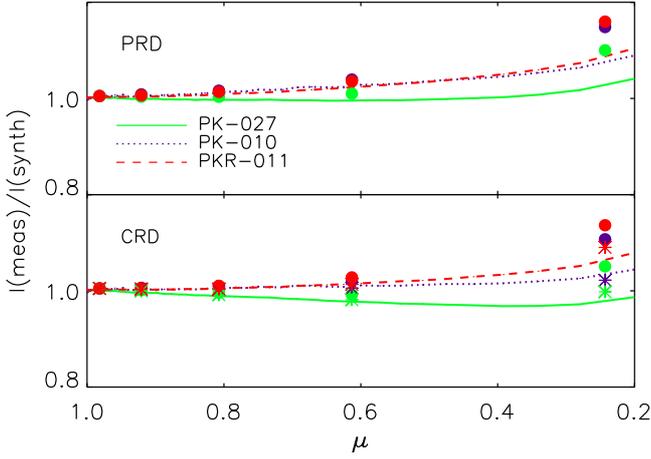}}
\caption{Ratio of the median of quiet Sun limb darkening measured 
in PSPT images to the quiet Sun limb darkening derived from the synthesis performed for the 
{\it PK-027} (green, solid line), {\it PK-010} (violet, 
dotted line),  and 
{\it PKR-011} (red, dashed line) filters by assuming either the PRD (top panel) or CRD (bottom panel) approximation. 
The lines indicate the results obtained by taking into account the CLV derived from the synthesis, while circles show the results 
derived by convolving the synthetic homogeneous quiet Sun atmosphere (associated with model B) by the model PSF. Asterisks show the ratio 
of  the CLVs measured in {\it PK} 
to the ones derived from contaminated inhomogeneous atmosphere 
composed of quiet Sun and network small features, associated with models B and D, respectively.   
}
\label{afig7} 
\end{figure}

We found that the
 convolution of the model PSF with synthetic homogeneous atmospheres 
leads to a significant decrease only in the intensity  and contrast values modeled for the brightest disk features  close to the solar limb. 
Figure \ref{afig7} shows that 
the discrepancy between  measured and 
modeled CLV for QS increases when moving toward the limb if straylight contamination of {\it PK} images is taken into account, except
when comparing the CLV  measured for {\it PK-027} and the one derived from CRD computations. Moreover, we found that {\it PK-027} measurements are in close agreement with 
the CLV obtained from the inhomogeneous atmospheres containing network features after their straylight contamination.

\begin{figure} 
\centering{\includegraphics[width=8.5cm]{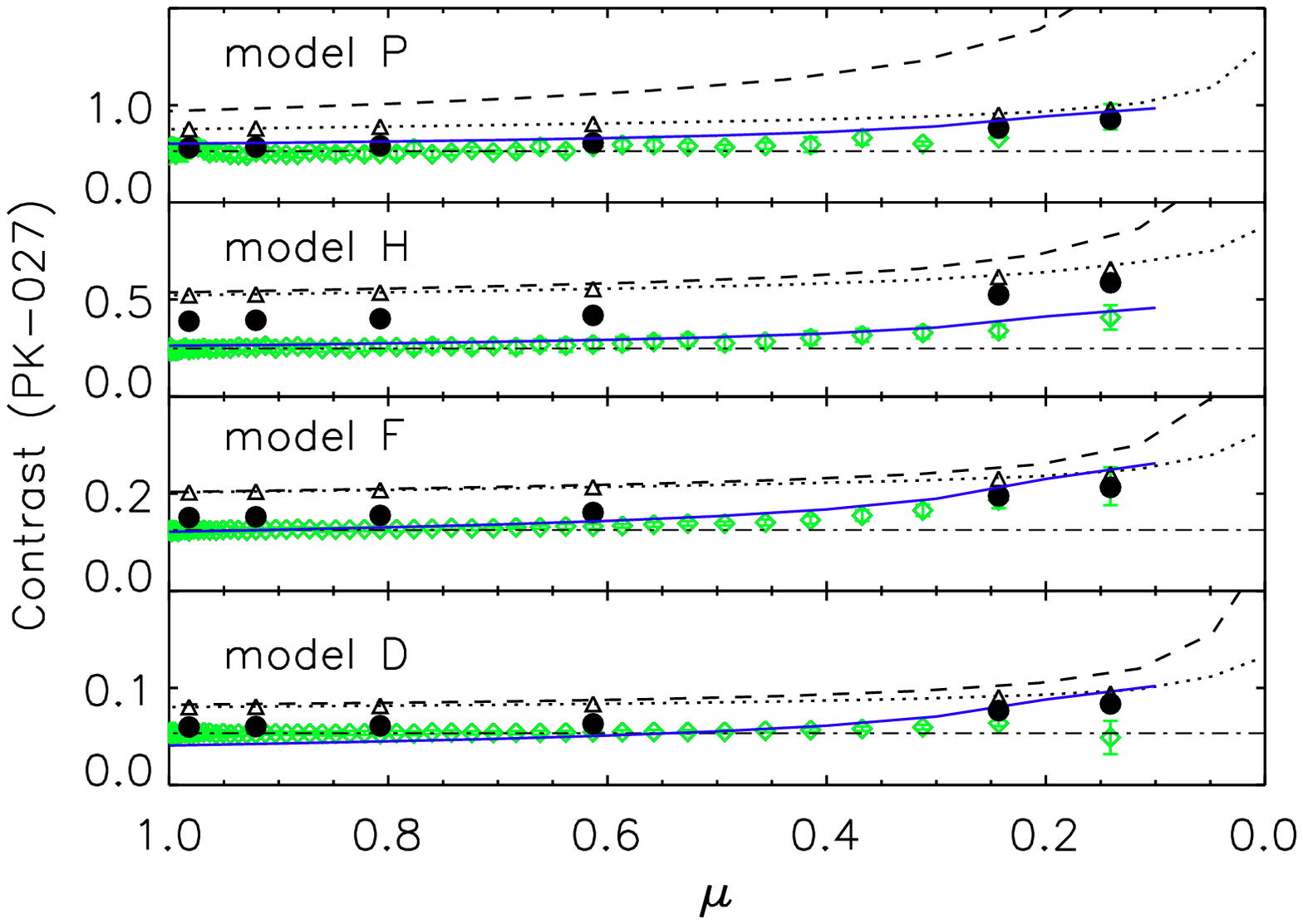}
\includegraphics[width=8.5cm]{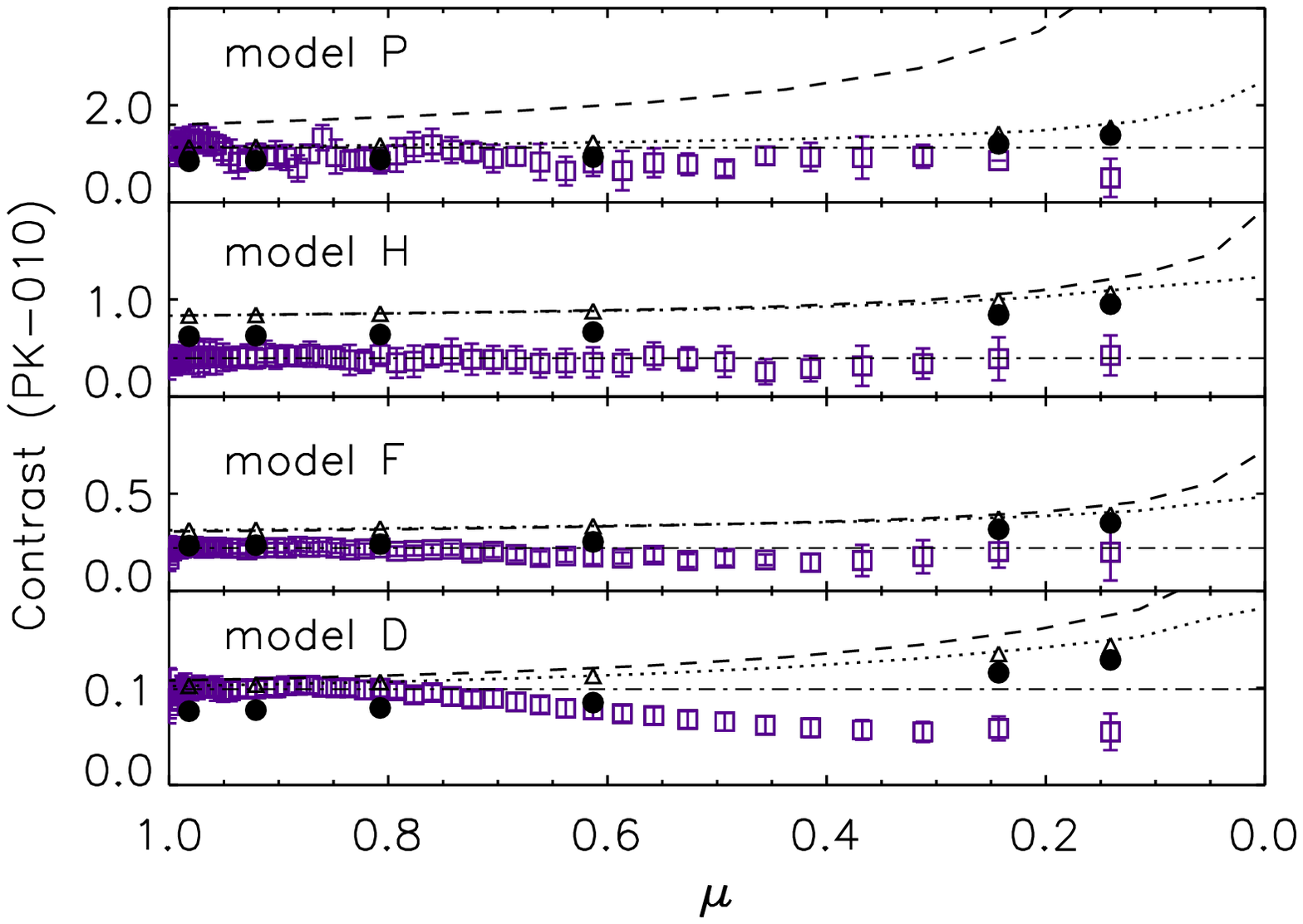}
\includegraphics[width=8.5cm]{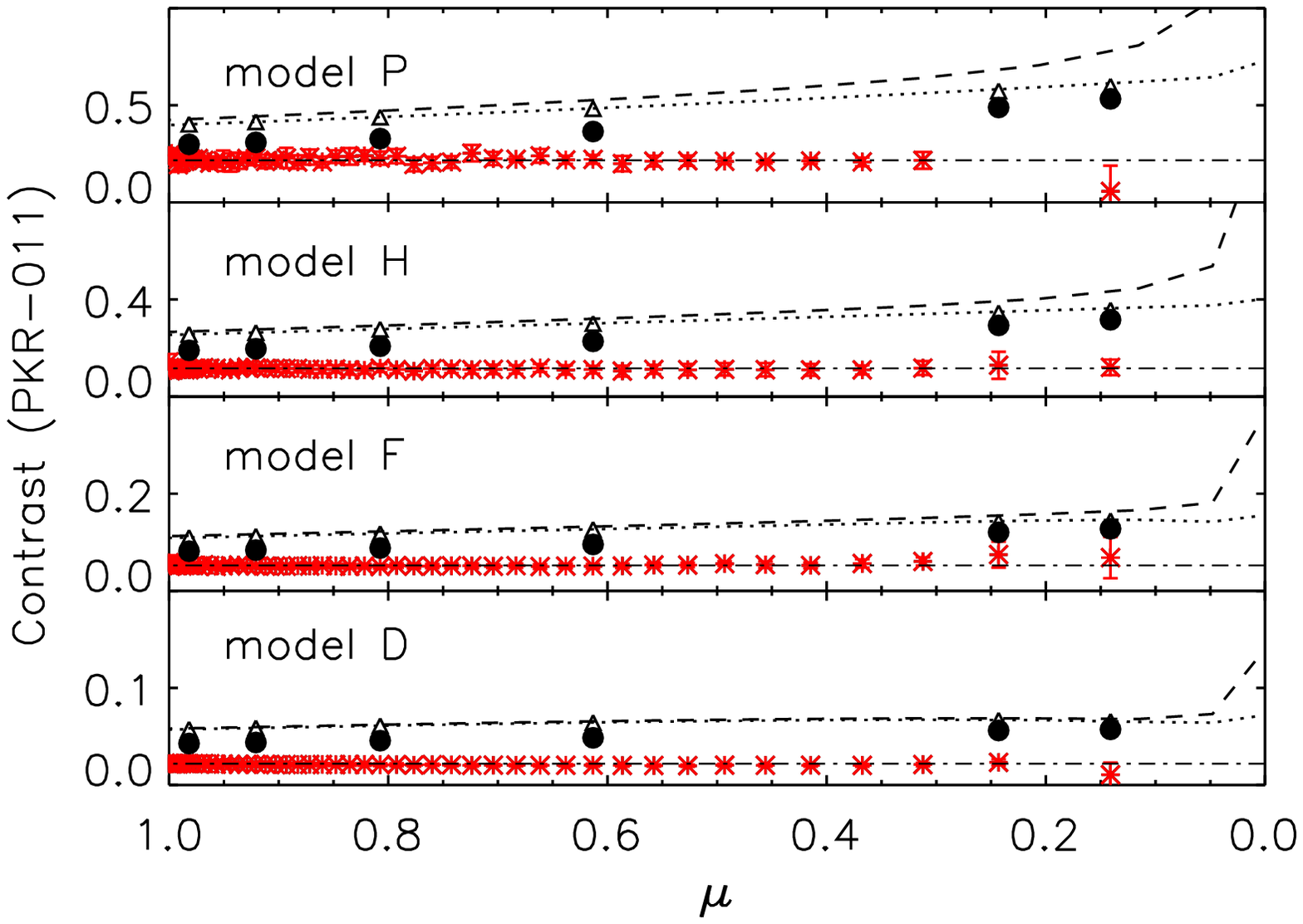}
}
\caption{CLV of median contrast values measured  (symbols) for various disk features identified in {\it PK-027} (top panel), {\it PK-010} (middle panel),   
and {\it PKR-011} (bottom panel) images un-restored for straylight degradation.  For each bandpass, the various sub-panels show 
measurement results for solar features ordered by decreasing contrast, i.e. from the top panel results for  
bright plage, plage, enhanced network, and network regions, respectively. The error bars represent the standard deviation of 
measurements. For each disk feature and bandpass, the 
dot-dashed line indicates 
the average of values measured at disk 
positions $\mu \ge$0.9, while dotted and dashed lines show the respective CLV derived from RH calculations with PRD and CRD and  
the reference model corresponding to the given feature, also indicated in the legend. 
Triangles and circles indicate  the results obtained after convolution of
 the synthetic atmospheres by the model PSF, 
for homogeneous and inhomogeneous atmospheres, respectively. 
In the top panel, the solid blue lines show the threshold values applied to the feature identification. 
}
\label{afig9} 
\end{figure}

\begin{figure}
\centering{\includegraphics[width=8.5cm]{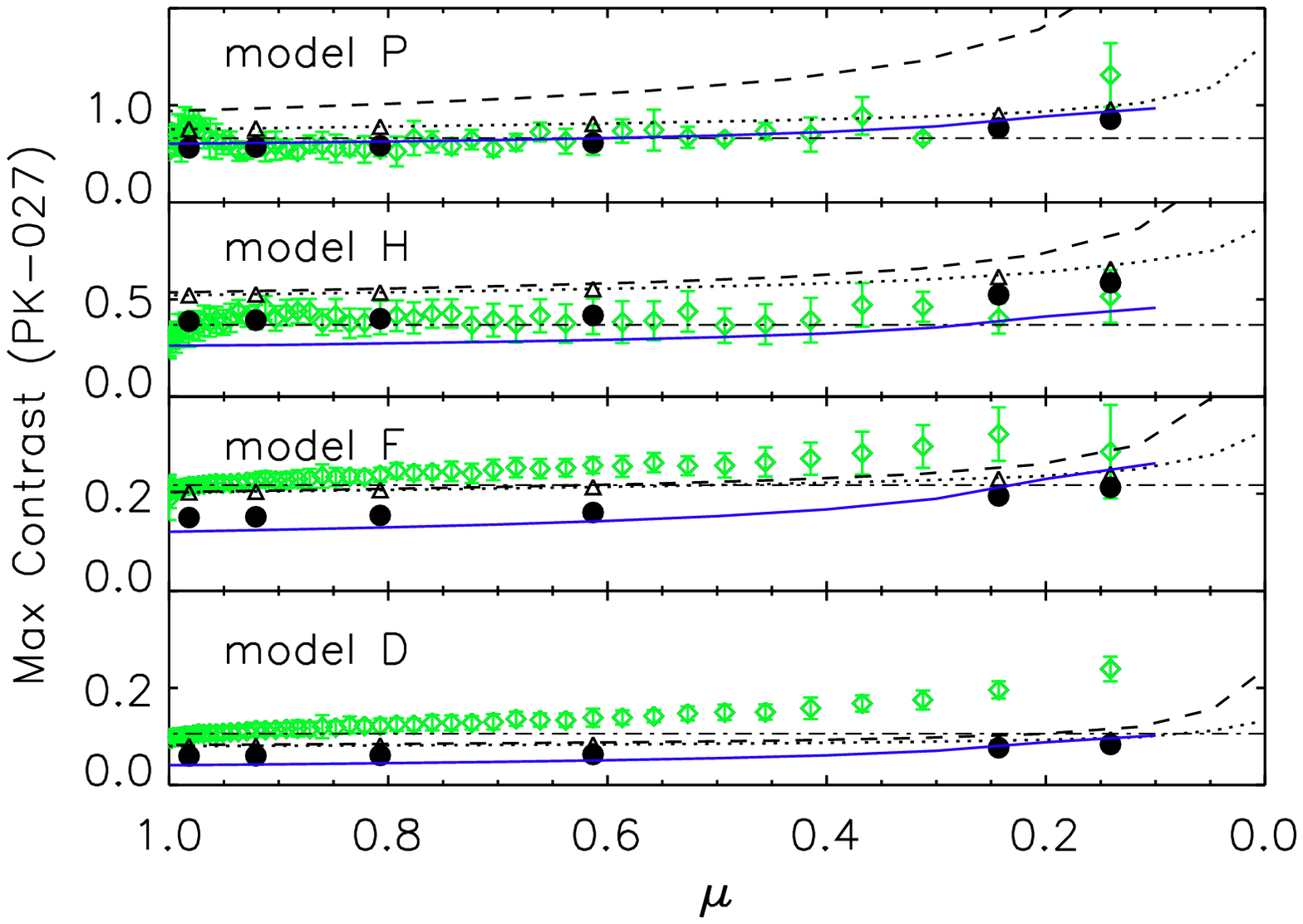}
\includegraphics[width=8.5cm]{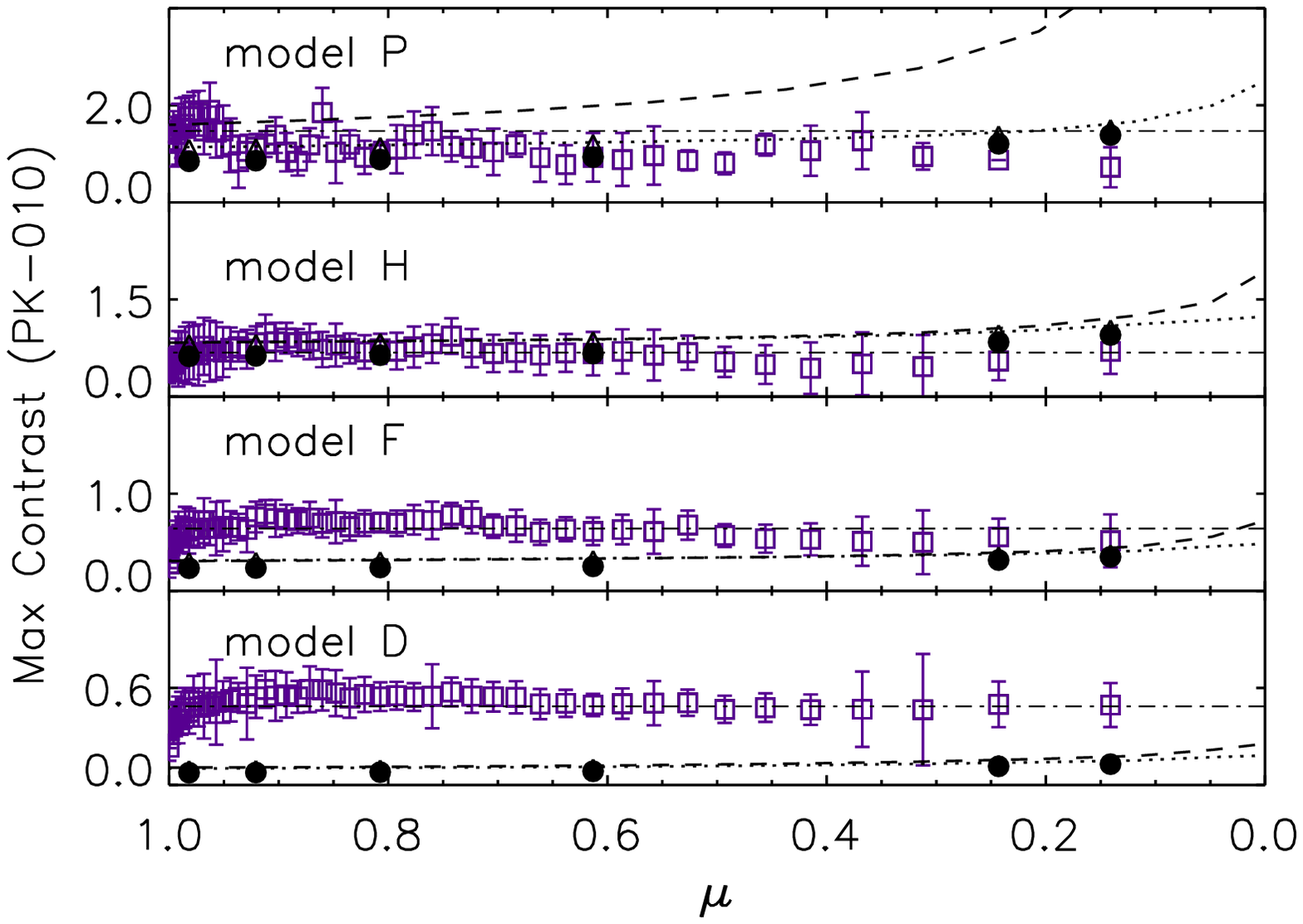}
\includegraphics[width=8.5cm]{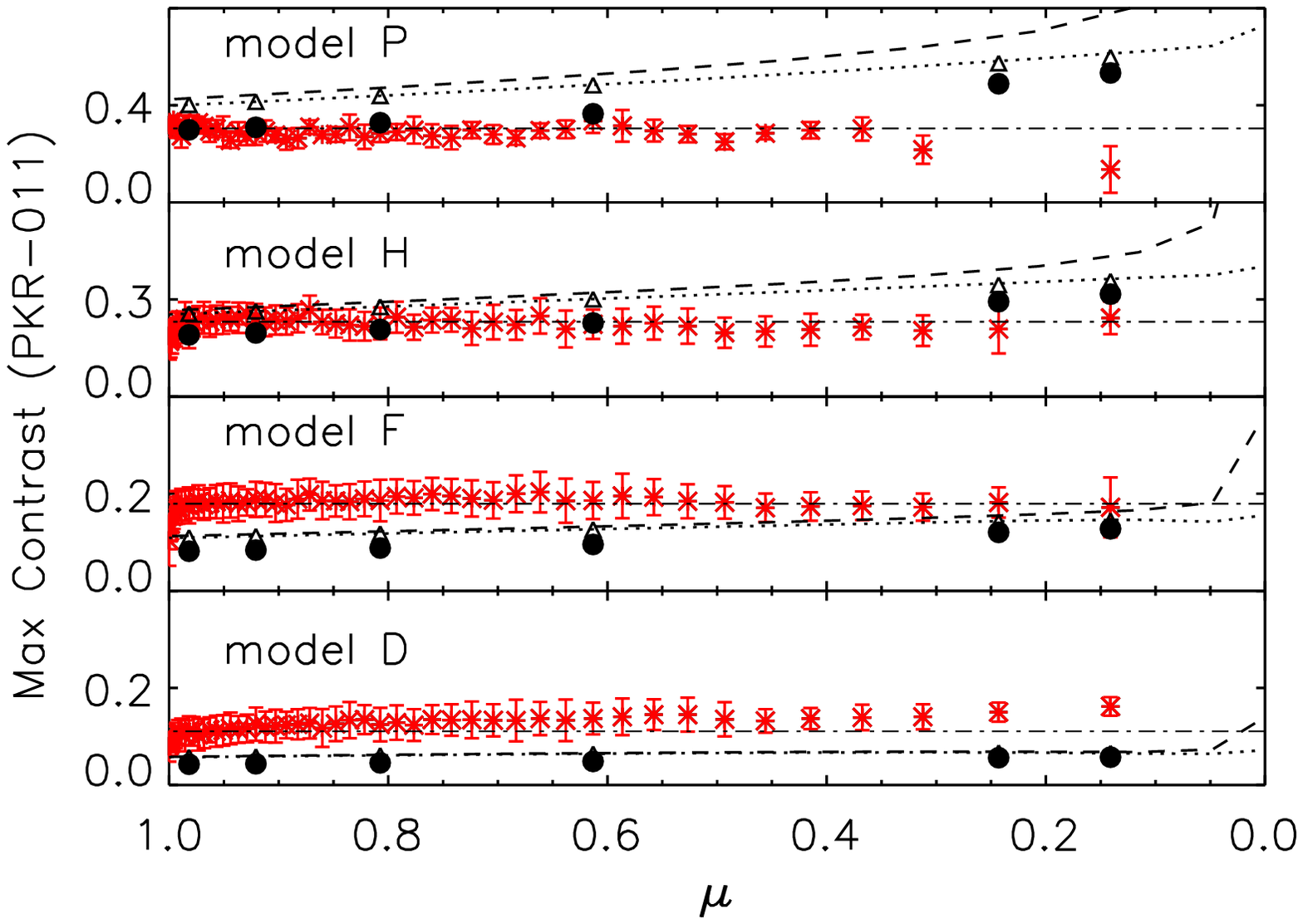}}
\caption{CLV of maximum contrast values measured  (symbols)
  for various disk features (model P to model D) in {\it PK-027} (top panel), {\it PK-010} 
(middle panel), and {\it PKR-011} (bottom panel) images un-restored for straylight degradation. Legend as given in Fig. \ref{afig9}. 
}
\label{afig9b} 
\end{figure}

On the other hand, we found that the convolution of synthetic inhomogeneous atmospheres with the same PSF leads to a significant decrease in 
 contrast values estimated for all the features and disk positions. 
This is illustrated by Figs. \ref{afig9} and \ref{afig9b}, which compare the results presented in Sect. 4, and the CLV derived from the 
convolution of the synthetic atmospheres with the model PSF. The contrast values estimated in the  synthetic atmospheres convolved by the PSF 
are by up to nearly a factor two lower than those obtained in the original RH computations. Therefore, these results suggest a closer general 
agreement between median contrast values 
 measured in {\it PK} and the contrast values derived from the RH synthesis than presented in Sect. 4, if straylight 
 contamination of small-scale disk features is taken into account.  

Our second test analysis was performed on {\it PK} images restored for straylight contamination by the deconvolution of 
original images 
with a model PSF derived from the analysed data. We applied the method and algorithm developed by \citet{criscuoli2008} 
 to the whole sample of {\it PK} good 
images analysed in Sect. 4.  

\begin{figure} 
\centering{\includegraphics[width=8.5cm]{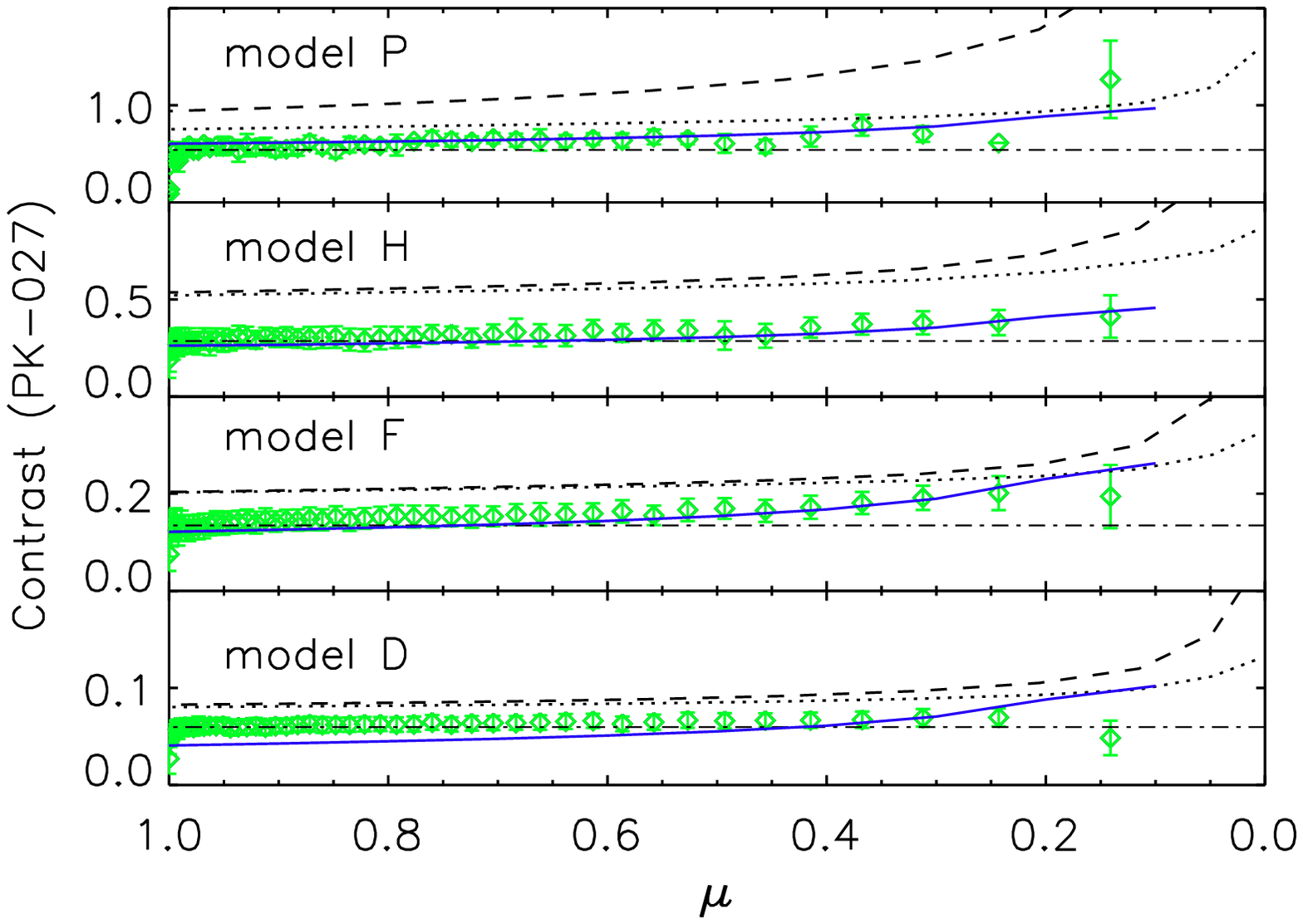}
\includegraphics[width=8.5cm]{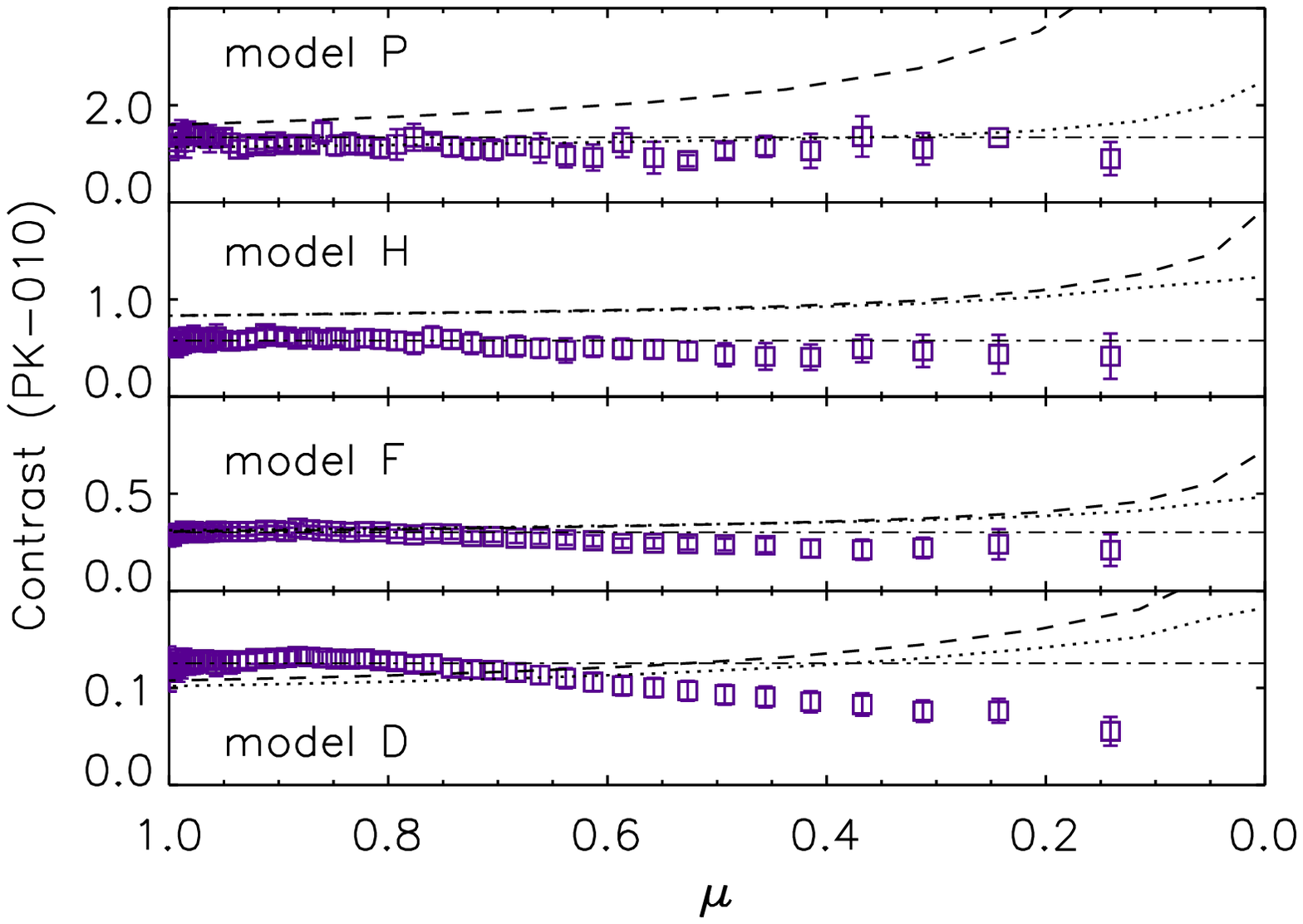}
\includegraphics[width=8.5cm]{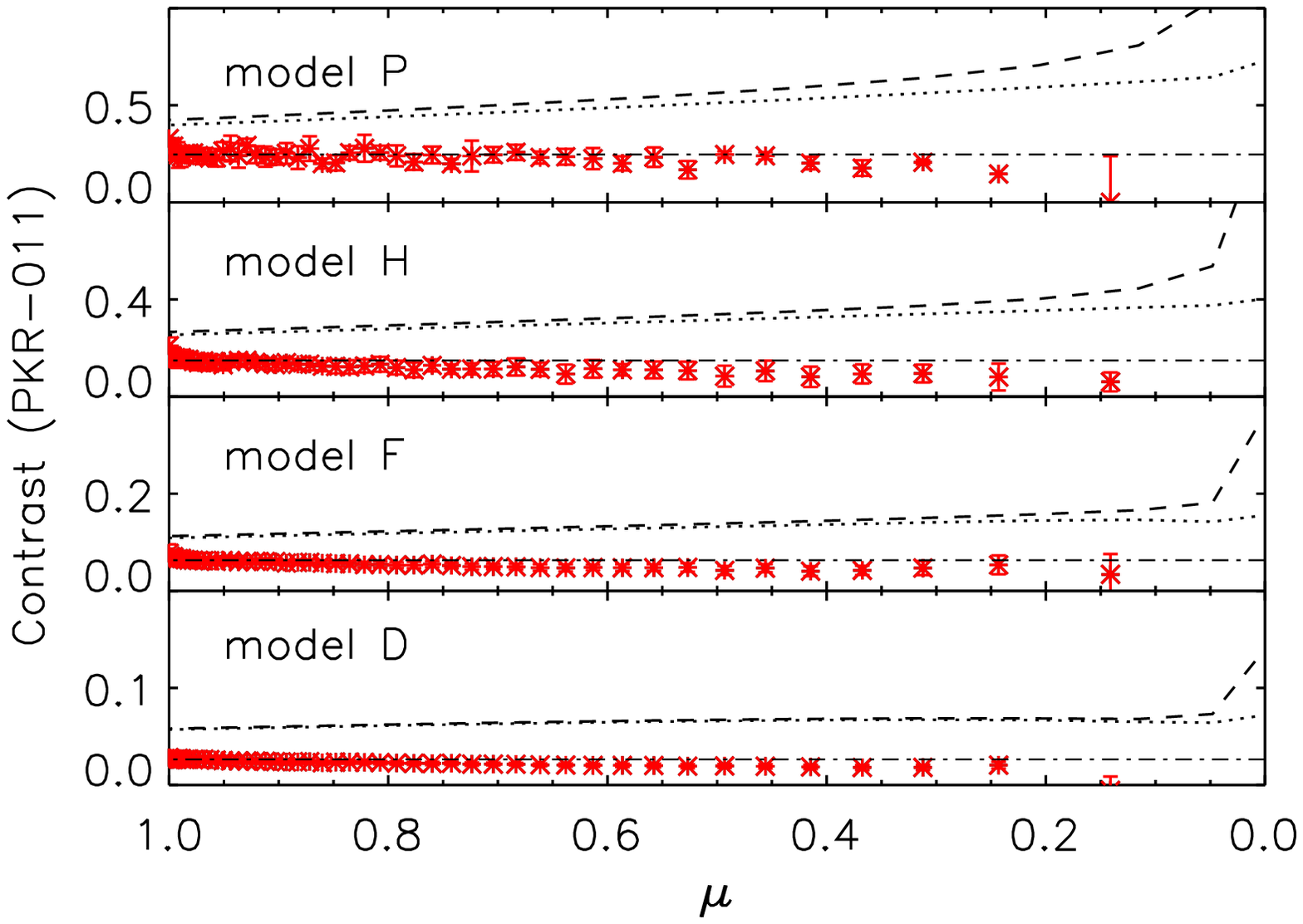}
}
\caption{CLV of median contrast values measured  (symbols) for various disk features identified in {\it PK-027} (top panel), {\it PK-010} (middle panel),   
and {\it PKR-011} (bottom panel) images restored from straylight degradation.  For each bandpass, the various sub-panels show 
measurement results for solar features ordered by decreasing contrast, i.e. from the top panel results for  
bright plage, plage, enhanced network, and network regions, respectively. The error bars represent the standard deviation 
of measurements. For each disk feature and bandpass, the 
dot-dashed line indicates 
the average of values measured at disk 
positions $\mu \ge$0.9, while dotted and dashed lines show the respective CLV derived from RH calculations with PRD and CRD and  
the reference model corresponding to the given feature, as also indicated in the legend. 
In the top panel, the solid blue lines show the threshold values applied to the feature identification. 
}
\label{afig8a} 
\end{figure}

\begin{figure}
\centering{\includegraphics[width=8.5cm]{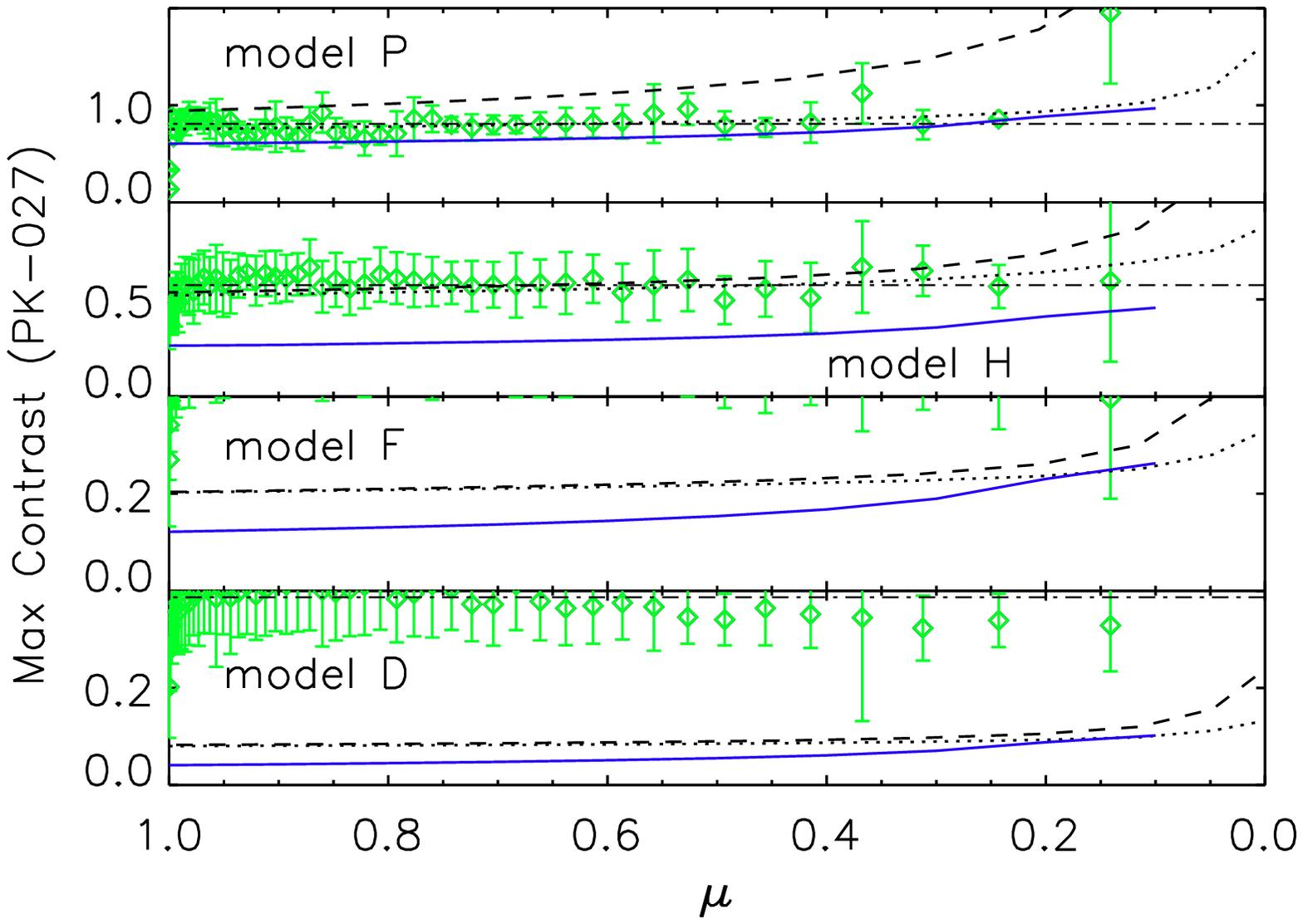}
\includegraphics[width=8.5cm]{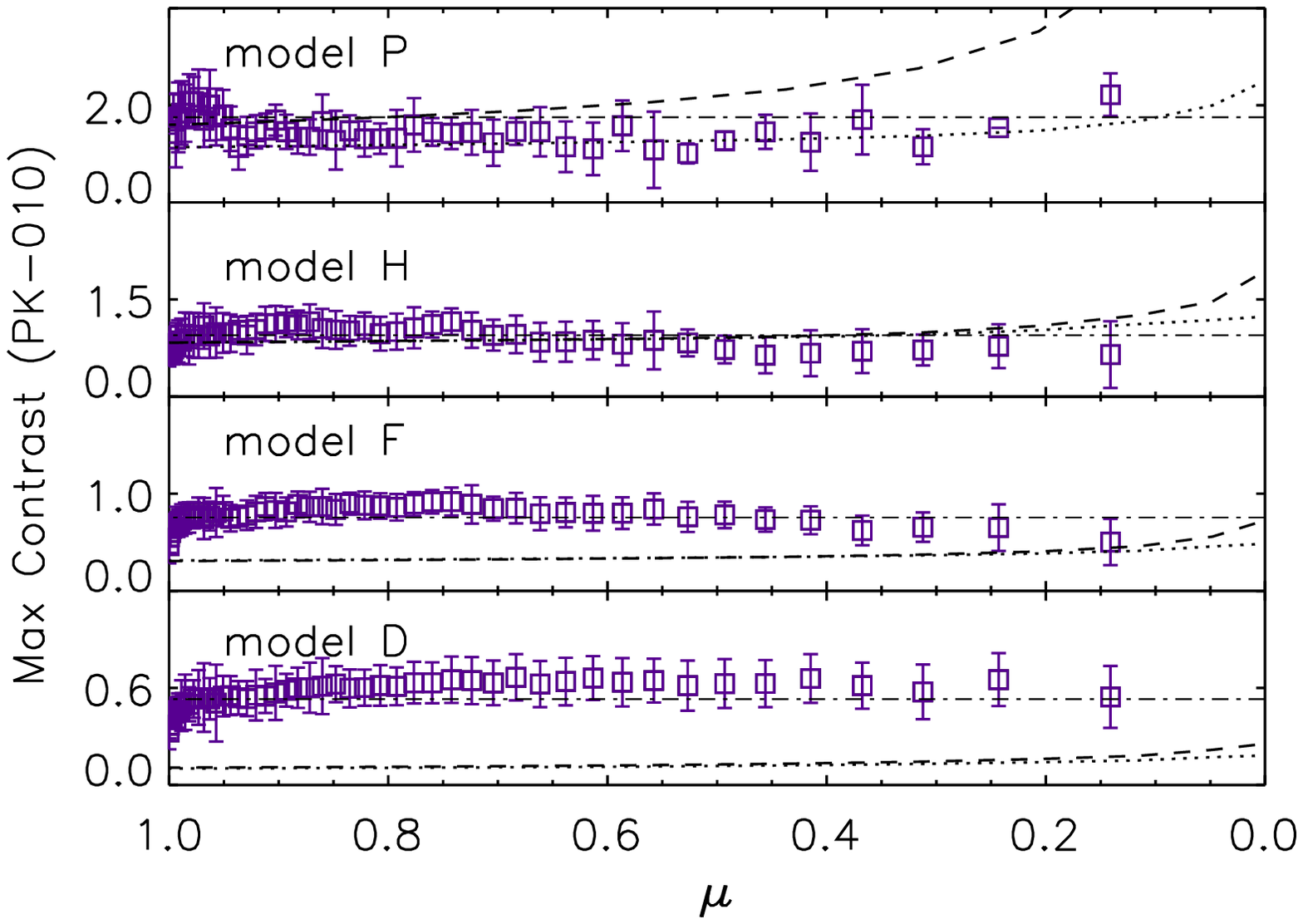}
\includegraphics[width=8.5cm]{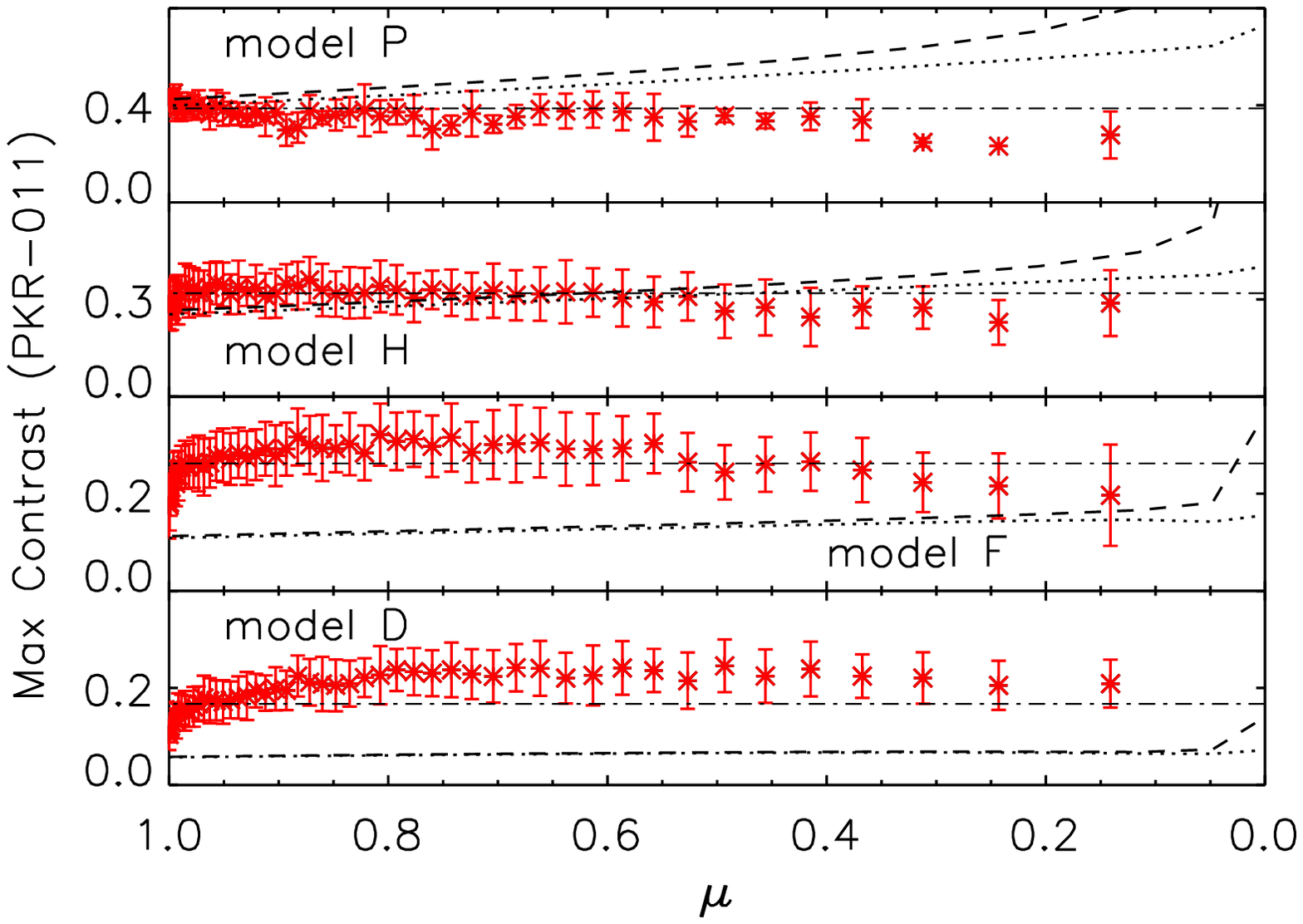}}
\caption{CLV of maximum contrast values measured  (symbols)
  for various disk features (model P to model D) in {\it PK-027} (top panel), {\it PK-010} 
(middle panel), and {\it PKR-011} (bottom panel) images restored from straylight degradation. 
For {\it PK-027}, measurement results for enhanced network and network features 
(associated with model F and D of Fontenla et al. (2009), respectively)
lie outside the plotted range, that is the range utilized for results from un-restored images. Legend as given in Fig. \ref{afig8a}. 
}
\label{afig8b} 
\end{figure}

Figures \ref{afig8a} and \ref{afig8b} compare model results presented in Sect. 4 and the CLV of contrast values 
measured for the various solar features on images corrected for straylight degradation. We found that image restoration leads 
to a slight increase in the average contrast 
values measured for all solar features with respect to the values  derived from un-restored images, in particular for the 
features identified in {\it PK-027} and {\it PK-010} images. In addition, we found an increase in the maximum 
contrast values measured for all the features, filter bandpasses, 
and disk positions, in particular for the network features in {\it PK-027} images. 
We also found  slightly closer agreement between model predictions and contrast values measured in restored images than derived from un-restored images. 
We found that    
 model predictions and contrast values measured at disk center ($\mu \ge 0.9$) in restored images indeed differ by a factor ranging from 
 0.81 to 2.17, 
 depending on the solar 
feature and filter bandpass, while  the same quantity evaluated on un-restored images ranges from 0.98 to 2.58. Modeled and measured values differ, 
on average, by a factor of 1.45 after taking into account straylight effects, compared to a factor of 1.71 for  measurements in 
un-restored images at the same disk positions.


\end{document}